\documentclass[prd,aps,twocolumn,floatfix]{revtex4}
\usepackage{amssymb,graphicx}
\usepackage{epsfig}
\usepackage[usenames]{color}
\usepackage{hyperref}
\usepackage{appendix}
\def\jnl@style{\it}
\def\aaref@jnl#1{{\jnl@style#1}}
\def\aaref@jnl#1{{\jnl@style#1}}

\def\aj{\aaref@jnl{AJ}}                   % Astronomical Journal
\def\araa{\aaref@jnl{ARA\&A}}             % Annual Review of Astron and Astrophys
\def\apj{\aaref@jnl{ApJ}}                 % Astrophysical Journal
\def\apjl{\aaref@jnl{ApJ}}                % Astrophysical Journal, Letters
\def\apjs{\aaref@jnl{ApJS}}               % Astrophysical Journal, Supplement
\def\ao{\aaref@jnl{Appl.~Opt.}}           % Applied Optics
\def\apss{\aaref@jnl{Ap\&SS}}             % Astrophysics and Space Science
\def\aap{\aaref@jnl{A\&A}}                % Astronomy and Astrophysics
\def\aapr{\aaref@jnl{A\&A~Rev.}}          % Astronomy and Astrophysics Reviews
\def\aaps{\aaref@jnl{A\&AS}}              % Astronomy and Astrophysics, Supplement
\def\azh{\aaref@jnl{AZh}}                 % Astronomicheskii Zhurnal
\def\baas{\aaref@jnl{BAAS}}               % Bulletin of the AAS
\def\jrasc{\aaref@jnl{JRASC}}             % Journal of the RAS of Canada
\def\memras{\aaref@jnl{MmRAS}}            % Memoirs of the RAS
\def\mnras{\aaref@jnl{MNRAS}}             % Monthly Notices of the RAS
\def\pra{\aaref@jnl{Phys.~Rev.~A}}        % Physical Review A: General Physics
\def\prb{\aaref@jnl{Phys.~Rev.~B}}        % Physical Review B: Solid State
\def\prc{\aaref@jnl{Phys.~Rev.~C}}        % Physical Review C
\def\prd{\aaref@jnl{Phys.~Rev.~D}}        % Physical Review D
\def\pre{\aaref@jnl{Phys.~Rev.~E}}        % Physical Review E
\def\prl{\aaref@jnl{Phys.~Rev.~Lett.}}    % Physical Review Letters
\def\pasp{\aaref@jnl{PASP}}               % Publications of the ASP
\def\pasj{\aaref@jnl{PASJ}}               % Publications of the ASJ
\def\qjras{\aaref@jnl{QJRAS}}             % Quarterly Journal of the RAS
\def\skytel{\aaref@jnl{S\&T}}             % Sky and Telescope
\def\solphys{\aaref@jnl{Sol.~Phys.}}      % Solar Physics
\def\sovast{\aaref@jnl{Soviet~Ast.}}      % Soviet Astronomy
\def\ssr{\aaref@jnl{Space~Sci.~Rev.}}     % Space Science Reviews
\def\zap{\aaref@jnl{ZAp}}                 % Zeitschrift fuer Astrophysik
\def\nat{\aaref@jnl{Nature}}              % Nature
\def\iaucirc{\aaref@jnl{IAU~Circ.}}       % IAU Cirulars
\def\aplett{\aaref@jnl{Astrophys.~Lett.}} % Astrophysics Letters
\def\apspr{\aaref@jnl{Astrophys.~Space~Phys.~Res.}}
\def\bain{\aaref@jnl{Bull.~Astron.~Inst.~Netherlands}} 
\def\fcp{\aaref@jnl{Fund.~Cosmic~Phys.}}  % Fundamental Cosmic Physics
\def\gca{\aaref@jnl{Geochim.~Cosmochim.~Acta}}   % Geochimica Cosmochimica Acta
\def\grl{\aaref@jnl{Geophys.~Res.~Lett.}} % Geophysics Research Letters
\def\jcp{\aaref@jnl{J.~Chem.~Phys.}}      % Journal of Chemical Physics
\def\jgr{\aaref@jnl{J.~Geophys.~Res.}}    % Journal of Geophysics Research
\def\jqsrt{\aaref@jnl{J.~Quant.~Spec.~Radiat.~Transf.}}
\def\memsai{\aaref@jnl{Mem.~Soc.~Astron.~Italiana}}
\def\nphysa{\aaref@jnl{Nucl.~Phys.~A}}   % Nuclear Physics A
\def\physrep{\aaref@jnl{Phys.~Rep.}}   % Physics Reports
\def\physscr{\aaref@jnl{Phys.~Scr}}   % Physica Scripta
\def\planss{\aaref@jnl{Planet.~Space~Sci.}}   % Planetary Space Science
\def\procspie{\aaref@jnl{Proc.~SPIE}}   % Proceedings of the SPIE

\begin{document}

\title{Linking electromagnetic and gravitational radiation in
coalescing binary neutron stars}

\author
{Carlos Palenzuela$^{1}$, Luis Lehner$^{2}$, Steven L. Liebling$^{3}$, Marcelo Ponce$^{4}$,  \\
Matthew Anderson$^{5}$,  David Neilsen$^{6}$,  and Patrick Motl$^{7}$
\\
\normalsize{$^{1}$Canadian Institute for Theoretical Astrophysics, Toronto, Ontario M5S 3H8,
 Canada}\\
\normalsize{$^{2}$Perimeter Institute for Theoretical Physics,Waterloo, Ontario N2L 2Y5, Canada}\\
\normalsize{$^{3}$Department of Physics, Long Island University, New York 11548, USA}\\
\normalsize{$^{4}$Department of Physics, University of Guelph, Guelph, Ontario N1G 2W1, Canada}\\
\normalsize{$^{5}$ Pervasive Technology Institute, Indiana University, Bloomington, IN 47405, USA}\\
\normalsize{$^{6}$Department of Physics and Astronomy, 
Brigham Young University, Provo, Utah 84602, USA}\\
\normalsize{$^{7}$Department of Science, Mathematics and Informatics,
Indiana University Kokomo, Kokomo, IN 46904, USA}}

\begin{abstract}
We expand on our study of the gravitational and electromagnetic emissions from
the late stage of an inspiraling neutron star binary as presented in Ref.~\cite{Palenzuela:2013hu}.
Interactions between the stellar magnetospheres, driven by the
extreme dynamics of the merger, can yield considerable outflows.
We study the gravitational and electromagnetic waves produced
during the inspiral and merger of a binary neutron star system using a
full relativistic, resistive MHD evolution code.
We show that the interaction between the stellar magnetospheres
extracts kinetic energy from the system and powers
radiative Poynting flux and heat dissipation. These features
depend strongly on the configuration of the initial  stellar magnetic moments.
Our results indicate that this power can strongly outshine
pulsars in binaries and have a distinctive angular and time-dependent pattern.
Our discussion provides more detail than Ref.~\cite{Palenzuela:2013hu}, showing
clear evidence of the different effects taking place during the inspiral.
Our simulations include a few milliseconds after the actual
merger and study the dynamics of the magnetic fields during the
formation of the hypermassive neutron star. We also briefly discuss
the possibility of observing such emissions.
\end{abstract}

\maketitle

%\tableofcontents

%%%%%%%%%%%%%%%%%%%%%%%%%%%%%%%%%%%%%%%%%%%%%%%%%%%%%%%%%%%%%%%
\section{Introduction}
%%%%%%%%%%%%%%%%%%%%%%%%%%%%%%%%%%%%%%%%%%%%%%%%%%%%%%%%%%%%%%%
Neutron stars are the most dense objects in the universe, anything 
with higher density must collapse to a black hole.
The neutron-rich matter in their cores
is compressed to very high densities, where the equation of state, and hence
their overall structure, is not completely understood.
The strong gravitational fields of these stars lead to interesting relativistic
effects, and their intrinsic magnetic fields are among the strongest arising in nature.
Single neutron stars serve as the inner engine of pulsars and power exciting
astrophysical events such as magnetar flares.
Interactions of a neutron star with
another neutron star or a black hole, likely play a fundamental role in the production of
gamma ray bursts. 

These compact binary systems are also among the most likely sources 
of detectable gravitational waves~(GW). Understanding their behavior is important 
for gravitational wave astronomy enabled by advanced detectors such as Advanced LIGO/VIRGO. 
Gravitational wave observations of neutron star binaries are expected
in the next few years, and the combination of these data with electromagnetic~(EM)
observations will provide
new opportunities to study the fundamental physics associated
with these stars.
For example, such combined observations could
reveal clues about their composition,
test strong-field gravity, and provide stringent constraints on current
models of their powerful electromagnetic emissions.

As an example of the important interplay between GW and EM data, consider the leading
model of short, hard gamma ray bursts~(SGRBs)
(see e.g.~\cite{2007PhR...442..166N} for a review). This model envisions
a merging binary system of either two neutron stars or a neutron star and a black hole as the key
ingredient to yield the energy and time scales observed in SGRBs.
If the initial binary consists of two neutron stars, the merger may result
in either the immediate collapse to a black hole or the creation of an
intermediate, hyper-massive neutron star, supported by thermal pressure and 
differential rotation, followed by a delayed collapse to a black hole. 
The interaction of the central compact object with 
a magnetic field can power radiation with the hard spectrum and short time 
scales characteristic of SGRBs.

Considerable effort has been devoted to understand
possible scenarios to explain these bursts concentrating on emission
after the merger. 
%Observations in
%both electromagnetic and gravitational wave bands will shed crucial light on this problem.
Recently however, there has also been significant interest in possible EM emissions
preceding collapse.  This interest has been 
motivated, in part,  by both the desire to maximize opportunities for gravitational
and electromagnetic wave detection, as well as surveys for EM transients that may be
precursors to sGRBs (e.g~\cite{2010ApJ...723.1711T}) and other
transients from compact binary mergers (e.g.~\cite{Kelley:2012tc,Kyutoku:2012fv}).
Precursor emissions may be generated by crust cracking due to
resonance effects~\cite{Tsang:2011ad} or magnetosphere 
interactions~\cite{1996A&A...312..937L,1996Vietri,2001MNRAS.322..695H,Piro:2012rq,Lai:2012qe}.
These studies rely on different approximations that may be adopted prior to
the last orbits of the system, when non-linear interactions and dynamics do
not overly complicate modeling possible emission mechanisms. 
It is in the final orbits, however, where nonlinear interactions are the 
strongest and the most powerful signals may be generated. As discussed 
recently in~\cite{Lehner:2011aa}, for example, the dynamics of the 
magnetosphere around a collapsing hypermassive 
neutron star can induce significant electromagnetic output.

When studying EM emission from dynamic neutron star systems, it is crucial 
to properly model the dynamics of the global electromagnetic field.
That is, incorporating in a consistent manner fields present in the stars
as well as their effects in the magnetosphere region surrounding the compact 
objects (see discussion in~\cite{Lehner:2011aa}). 
We do so here considering general relativity coupled to relativistic, 
resistive magnetohydrodynamics and study magnetized binary neutron stars,
monitoring closely the radiation produced by the system.

In this work we concentrate on the late inspiral and merger epoch
of a binary neutron star system and study possible electromagnetic and gravitational energy flux produced.  
Of particular interest is the localization of the  dissipative regions 
and the analysis of the topology of the resulting electromagnetic field.

This work is organized as follows, section II describes details of
our approach. We include in section III a discussion of models
presented to explain possible pre-merger emissions.  In Section IV
we describe results from numerical studies of three relevant systems,
discuss their main features and compar the obtained behavior. We
conclude in section V with final comments and discussions.

%%%%%%%%%%%%%%%%%%%%%%%%%%%%%%%%%%%%%%%%%%%%%%%%%%%%%%%%%%%%%%%
\section{Approach}
%%%%%%%%%%%%%%%%%%%%%%%%%%%%%%%%%%%%%%%%%%%%%%%%%%%%%%%%%%%%%%%

Our primary goal is to investigate how the strongly gravitating 
and highly dynamical behavior of a binary neutron star system can affect the plasma
in the magnetosphere such
that powerful electromagnetic emissions can be induced.
Such systems are natural candidates for loud gravitational wave emissions,
and the interplay of strong/highly dynamical gravity
with global electromagnetic fields should lead to bright electromagnetic signals.
To study this problem we exploit a
recently introduced framework incorporating general relativity and relativistic,
resistive magnetohydrodynamics~\cite{2013MNRAS.431.1853P} (which builds
from previous works ~\cite{Palenzuela:2008sf,Lehner:2011aa,2012arXiv1208.3487D})
to study the behavior of magnetically dominated plasma surrounding a binary
magnetized neutron star system.

In this approach, the full Einstein-Maxwell-hydrodynamic equations are employed
to model strongly gravitating compact stars and the effects of a global electromagnetic
field. Inside the star, the magnetic field is modeled within the ideal MHD limit. 
The conductivity is prescribed so that the resistive scheme smoothly transitions from the
ideal limit to the force-free limit outside the stars.
This transition is achieved by setting the conductivity dependent on the 
fluid density, such that the conductivity varies by several orders of magnitude.

To incorporate gravitational effects in complete generality, we adopt a 
BSSN formulation~\cite{Baumgarte:1998te,Shibata:1995we} of the Einstein equations as described in~\cite{Neilsen:2010ax}.
We use finite difference techniques on a regular, Cartesian grid to discretize
the system~\cite{SBP2,SBP3}. The geometric fields are discretized
with a fourth order accurate scheme satisfying the summation by
parts rule, while High Resolution Shock Capturing methods based on the HLLE
flux formulae with PPM reconstruction are used to discretize the fluid and
the electromagnetic variables~\cite{Anderson:2006ay,Anderson:2007kz}.

The time evolution of the resulting equations must address the appearance of certain
stiff terms arising from the resistive MHD scheme in (some of) the equations of motion.
Such terms are efficiently handled with an IMEX
(implicit-explicit) Runge-Kutta scheme, as described in~\cite{Palenzuela:2008sf,
2012arXiv1208.3487D,2013MNRAS.431.1853P}.
The explicit part of the time evolution is performed through the method of lines using
a third order accurate Runge-Kutta integration scheme, which helps to ensure
stability of the numerical implementation~\cite{Palenzuela:2008sf}.

To ensure sufficient resolution in an efficient manner, we employ adaptive mesh
refinement (AMR) via the HAD computational infrastructure that provides distributed,
Berger-Oliger style AMR~\cite{had_webpage,Liebling} with full sub-cycling
in time, together with an improved treatment of artificial
boundaries~\cite{Lehner:2005vc}. The refinement regions are determined using truncation
error estimation provided by a shadow hierarchy~\cite{Pretoriusphd} which adapts
dynamically to ensure the estimated error is bounded within a pre-specified tolerance.

%%%%%%%%%%%%%%%%%%%%%%%%%%%%%%%%%%%%%%%%%%%%%%%%%%%%%%%%%%%%%%%%%%%%%%%
\section{Preliminary Luminosity Estimates}
\label{sec:unipolar}
%%%%%%%%%%%%%%%%%%%%%%%%%%%%%%%%%%%%%%%%%%%%%%%%%%%%%%%%%%%%%%%%%%%%%%%

The understanding of possible electromagnetic precursors driven by a compact binary
system is an active area of research. In recent years several mechanisms have been
discussed in this context, relying on simplified models, to obtain relevant estimates.
A basic question here is what possible mechanisms
could yield sufficiently strong electromagnetic emissions to be detected by different
facilities (in suitable bands). Moreover, such detection might be further
aided by gravitational wave input which in future years will provide timing and
sky localization (in addition to other physical parameters).
The information provided by future GW observations may also lead to deeper and longer
investigations of signals on both fronts~(e.g.~\cite{Branchesi:2011mi,Kelley:2012tc,
lrr-2009-2,Bloom:2009vx}).

For the particular case of precursor signals from binary mergers the challenge is
to identify appropriate mechanisms that could act prior to the stars coming into contact and
yield significant (i.e., possibly observable) emissions. A few mechanisms have been
recently discussed. One of these is the emission of flares induced by
{\em resonant excitations of NS modes by tides}, which could induce
crust-cracking~\cite{Tsang:2011ad} and the consequent release of $10^{46-47}$ergs of
energy a few seconds before merger. Another relies on {\em unipolar induction},
which bears direct relevance to our present discussion and, for this reason,
is discussed in some detail next. (A related model~\cite{2012arXiv1212.0333M}, examines
possible emissions through Fermi acceleration at shocks created by Poynting flux-driven bubbles.)

The main mechanism in the unipolar inductor model is the extraction of stellar
kinetic energy by the interaction of the stellar magnetosphere with an external magnetic field.
In this case, one can consider each star to be interacting with the field of its companion.
The energy released will either reach its maximum just prior to merger or give rise to
episodic emissions depending on the resistance  of an assumed `effective' circuit from one
star to the other. Different models build upon this
possibility~\cite{2001MNRAS.322..695H,Piro:2012rq,Lai:2012qe}, predicting a transient
preceding the merger by (of order) a few seconds, possibly accompanied by a radio signal.

A good starting point for this discussion is the model considered by Ref.~\cite{Lai:2012qe}:
a binary system with a magnetized primary star and an unmagnetized secondary.
An estimate of the luminosity for such a configuration is given by
\begin{eqnarray}\label{Hansen_Lyutikov}
   {\cal L} &\approx& \xi_{\phi} v_{\rm{rel}} \frac{B_*^2 R_*^6 R_c^2}{2 a^6} \\
\label{Hansen_Lyutikov2}
     &\approx& %1 \times
                10^{41}   \xi_{\phi} \left( \frac{B_*}{10^{11} \rm G} \right)^2 \! %\,
                          \left( \frac{R_c}{13.6 \rm km}\right)^2 \! %\,
                          \left( \frac{a}{30 \rm km} \right)^{-13/2} \! {\rm ergs/s}
\end{eqnarray}
where $\xi_{\phi}=16  v_{\rm{rel}} R_{\rm{tot}}$ is the azimuthal twist
of flux tubes connecting both stars, $v_{\rm{rel}}$ is the relative velocity,
$R_{\rm{tot}}$ is the total resistance of the
circuit, $B_*,R_*$ are the field strength and radius of the primary star; $B_c,R_c$ the
field strength (assumed much weaker) and radius of the companion star and $a$ is the orbital
separation [notice that to obtain Eq.~(\ref{Hansen_Lyutikov2}) we assume
values $R_*=R_c=13.6$ km and $M_*=M_c=1.4 M_{\odot}$ ].
A key unknown here is the azimuthal twist which depends on the total resistance. This 
resistance, in turn, depends on the conductivity of the stars and on the properties
of the magnetosphere in between them. 
In the case of free space, $R_{\rm{tot}}=4\pi$  and with this choice the luminosity
becomes~\cite{2001MNRAS.322..695H,Lai:2012qe}
\begin{eqnarray}\label{Hansen_Lyutikov3}
   {\cal L} &\approx& 3 {\times} 10^{40}\! \left( \frac{B_*}{10^{11}\rm G} \right)^2 \! \left( \frac{R_c}{13.6 \rm km}\right)^2 
                         \! \left( \frac{a}{30 \rm km} \right)^{-7} \! \! \! \! {\rm ergs/s}
\end{eqnarray}
which can be re-expressed in terms of the orbital frequency $\Omega$ (assuming a Keplerian relation)
as
\begin{equation}\label{eq:fourteenthirds}
 {\cal L}  \approx 10^{41}\, \left( \frac{B_*}{10^{11}\rm G} \right)^2 
\left (\frac{\Omega}{\Omega_{\rm ISCO}} \right)^{14/3}~{\rm ergs/s}
\end{equation}
in terms of the fiducial angular frequency $\Omega_{\rm ISCO}=4758$~rad/s chosen to be that 
of a particle at the inner-most, stable,
circular orbit for a non-spinning black hole of mass $2.9 M_{\odot}$ (this frequency is a good mark of the
onset of the  plunging behavior~\cite{Baiotti:2011am,Buonanno:2007sv}). 
Notice this luminosity is well below the saturation point of plasma acceleration,
which happens when the plasma energy density becomes comparable to the
energy density of the magnetic field companion~\cite{1996Vietri}; thus the
magnetosphere remains present. 

The above estimate treats this problem as essentially a quasi-adiabatic process with
only one star dominating the magnetic effects, so that the induced circuit can be analyzed in simple terms.
However more complex behavior arises when both stars are magnetized. For instance, already
at the quasi-adiabatic level
when the companion is weakly magnetized, the radius at which induction occurs is not
$R_c$ but instead an effective radius that depends, at least, on the relative magnetizations of the stars.

For instance, in the simplest case of aligned dipolar fields the companion's field will shield the 
effects of the primary at some effective radius, $R_{\rm eff}$.  We can estimate this radius
by assuming equality of the magnetic field produced by each star at that location in terms of their
magnetic moments
\begin{equation}
\mu_* \left(a - R_{\rm eff} \right)^3 =   \mu_c R_{\rm eff}^3
\end{equation}
where $\mu_*$ is the moment of the primary and $\mu_c$ is the moment of the companion.
Because each star is essentially perfectly conducting with frozen magnetic flux in its interior,
we have a lower bound $R_{\rm eff} \ge R_c$.
Assuming $a \gg R_{\rm eff}$, $B_c\propto \mu_c$ and $B_*\propto \mu_*$, we then have 
\begin{equation}
R_{\rm eff} = {\rm max}\left( a \, \left( \frac{B_c}{B_*} \right)^{1/3}, R_c \right).
\end{equation}

The luminosity of Eq.~(\ref{Hansen_Lyutikov3}) with $R_{\rm eff} \propto \left(B_c/B_*\right)^{1/3}$ yields
\begin{eqnarray}\label{Hansen_Lyutikov4}
   {\cal L} &\sim& 1.5 \times 10^{41}\, \left( \frac{B_*}{10^{11}\rm G} \right)^2 \, \left( \frac{B_c}{B_*}\right)^{2/3} 
                         \, \left( \frac{a}{30 \rm km} \right)^{-5} {\rm ergs/s}
\end{eqnarray}
or, in terms of the orbital frequency,
\begin{equation} \label{ui_effect}
 {\cal L} \sim 2 \times 10^{41}\, \left( \frac{B_*}{10^{11}\rm G} \right)^2 \, \left( \frac{B_c}{B_*}\right)^{2/3} \left( \frac{\Omega}{\Omega_{\rm ISCO}}\right)^{10/3} ~\rm ergs/s  .
\end{equation}
This estimate already indicates that the relative magnetization of the stars introduces departures from the basic
unipolar result of Eq.~(\ref{Hansen_Lyutikov3}).
One naturally expects further departures due to the dynamics of the magnetospheres;
in particular,
significant reconnections of the magnetic field lines may be induced which would
depend on the orientation of the magnetic moments. In addition, as the stars approach merger,
the increasing strength of the gravitational potential,
the rapidly changing geometry of the spacetime, and the stellar dynamics will all affect
the magnetic field. 
Naturally, estimating these effects is difficult, at best.  Instead, we proceed with
numerical solutions to unravel the possible phenomenology. 

%%%%%%%%%%%%%%%%%%%%%%%%%%%%%%%%%%%%%%%%%%%%%%%%%%%%%%%%%%%%%%%%%%%%%%%
\section{Binary Neutron Star simulations}
%%%%%%%%%%%%%%%%%%%%%%%%%%%%%%%%%%%%%%%%%%%%%%%%%%%%%%%%%%%%%%%%%%%%%%%

We consider the late orbiting behavior and merger of  magnetized binary neutron
star systems. The magnetic field dynamics within the stars and the gravitational
wave output from these systems have previously been studied through numerical simulations
\cite{2008PhRvL.100s1101A,Liu:2008xy,Giacomazzo:2009mp,Rezzolla:2011da}.

Our primary goal here is to understand magnetic effects arising
in the magnetosphere and the dependence of these effects upon the initial magnetic configurations
of the stars. Because tidal effects play a relatively minor role (and only close to the
merger epoch)~\cite{Read:2009yp,Damour:2012yf}, 
and because electromagnetic interactions do not influence the dynamics of the binary for realistic field strengths~\cite{Ioka:2000yb},
our studies are readily applicable to generic binary systems. %, modulo field topologies. 
For simplicity, we concentrate on a binary initially described by a pair of identical, 
irrotational neutron stars in a quasicircular orbit with an initial separation 
$a=45 {\rm km}$,
$\Omega_o=1.85 {\rm rad}/{\rm ms}$.

The initial geometric and matter configurations for this system are obtained with the
LORENE library~\cite{lorene}, which adopts a polytropic equation of state
$P=K \rho^{\Gamma}$ with $\Gamma=2$ and $K=123$, approximating  cold nuclear matter. 
During the evolution,
the fluid is modeled as a magnetized perfect fluid with an ideal gas
equation of state that allows for shocks. Note that the dynamics and interactions of the 
electromagnetic (e.g.~\cite{Ioka:2000yb,2008PhRvL.100s1101A,Lai:2012qe}) and
gravitational (e.g.~\cite{Read:2009yp,Baiotti:2011am}) fields
are largely insensitive to the choice of equation of state during the inspiral.

For convenience, unless otherwise noted, we adopt geometrized units $G=c=1$,
such that all quantities, including mass ($M$) and time ($T$) have units of length ($L$).
Additionally we set the solar mass $M_\odot \equiv 1$. The above choice then constitutes
our ``code units,'' and the relation between code and physical length units is given by the 
multiplicative factor $1.48 \, \rm{km}$. Unless otherwise noted by their appropriate 
physical units, we will be displaying code units in our figures.

We adopt individual stars having
a baryonic mass $M=1.62 M_{\odot}$, radius $R_*=13.6 {\rm km}$, and a
magnetic moment $\mu_i$ that describes a dipolar magnetic field $B^i$ in
the comoving frame of each star.  The magnetic moment is aligned with
the orbital angular momentum, i.e. $\mu_i=(0,0,\mu)$. The non-trivial component
is related to the radial
magnetic field at the pole of the star, $B_*$, by the relation $\mu = B_* R_*^3$.
In our simulations we choose $B_* = 1.5 \times 10^{11} {\rm G}$, 
a value which is relatively high for neutron stars in binaries but still realistic.
The electric field is obtained from the
ideal MHD condition ${\bf E} = - {\bf v} \times {\bf B}$, where the
velocity in the star is given by the orbital motion and we 
assume  that the magnetosphere is initially at rest.

To cover the range of possible cases and to gain insight into the 
underlying behavior,
we consider three different initial configurations of magnetic moments of
each star~$(i)$:
\begin{itemize}
\item $U/U$ : aligned case $\mu_{(1)}=\mu_{(2)}=\mu$,
\item $U/D$ : anti-aligned case  $\mu_{(1)}=-\mu_{(2)}=\mu$,
\item $U/u$ : one-dominant, aligned case  $\mu_{(1)}=100\, \mu_{(2)}=\mu$.
\end{itemize}
Notice that the last case~($U/u$) has magnetic moments similar to those estimated in
the double binary pulsar PSR J0737-3039~\cite{2003Natur.426..531B}, and the
orientation of the moments resembles one of two configurations obtained in a model
of that same system~\cite{2004Natur.428..919J}. Future work will explore
configurations with other inclinations of the magnetic moments with respect to the orbital angular momentum.

Our numerical domain extends up to $L=320~{\rm km}$ and contains five,
centered FMR grids with decreasing side-lengths (and twice as well resolved)
such that the highest resolution grid has $\Delta x = 300~{\rm m}$
and extends up to $58~{\rm km}$, covering both stars and the inner
part of the magnetosphere. We have computed the Poynting-vector
luminosity at three different surfaces, the furthest located at
$R_{\rm ext}= 180~{\rm km}$. Within this setup, we have evolved the three
cases described above and have studied the resulting behavior.
We have also
compared coarser solutions obtained for the $U/U$ and $U/D$ cases
and have found
that the qualitative features of the magnetic fields are very
similar and that the luminosity differs only by a few percent.
Our main results are summarized in the following sections.

%%%%%%%%%%%%%%%%%%%%%%%%%%%%%%%%%%%%%%%%%%%%%%%%%%%%%%%%%%%%%%%
\subsection{Orbital motion and gravitational waves}
\label{subsec:GravResults}

In all cases, the field strengths considered have a negligible effect  in 
the orbital dynamics of the system up to the merger~\cite{Ioka:2000yb,
2008PhRvL.100s1101A}.
Consequently, the  three cases studied progress to merger in exactly the same
way, producing the same gravitational signal.

\begin{figure}
\begin{center}
	\epsfig{file=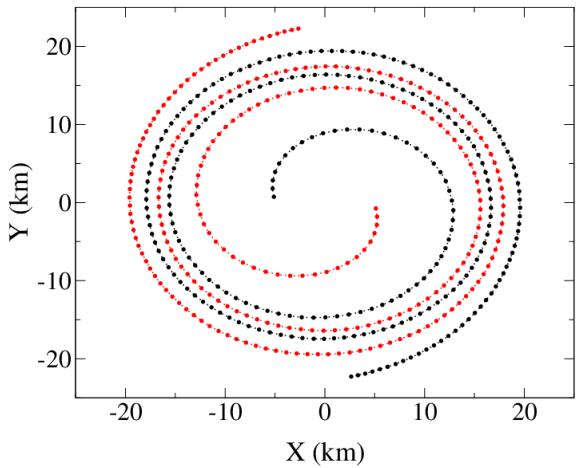,width=0.75\columnwidth,angle=-90}
	\caption{Trajectory of the binary as measured  by the location of the maximum densities as functions of time.
The system undergoes about 2.5 orbits before the stars come into contact.
   }
	\label{fig:trajectories}
\end{center}
\end{figure}

\begin{figure}
\begin{center}
        \epsfig{file=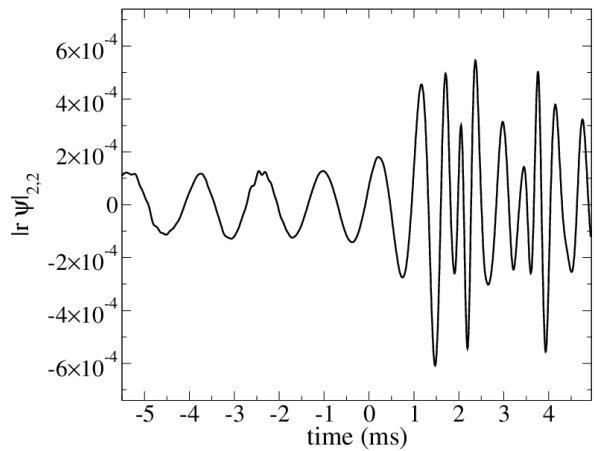, width=0.7\columnwidth,angle=-90}
%	\epsfig{file=./figures/PN/GW_lum-PN_numtraj.ps,height=2.5in,width=1.5in,angle=-90}\\
%	\caption{(Left panel) Gravitational waves, $c_{22}$ and luminosity; extracted in a 
%``hybrid-PN'' approach employing the numerical trajectories. (Right panel) Gravitational
%wave luminosity vs time during the time studied. Near the merger epoch the luminosity
%is $\approx 2-3 \times 10^{55}$ ergs/s {\bf MARCELO let's check}
	\caption{The dominant $l=m=2$ mode of the gravitational wave
extracted at $R_{\rm ext}=180 {\rm km}$.  % SLL (already said in text) and shifted in time so that
% the peak of the luminosity (i.e., the merger) happens at $t=0$.
The time $t=0$ is set as the moment when the stars first make contact.
   }
	\label{fig:gravwaves}
\end{center}
\end{figure}

Fig.~\ref{fig:trajectories} 
illustrates the path of the stars by displaying the location of the maximum of
the fluid densities. The stars orbit about
each other for $\approx 2.5$ orbits before they come into contact. 
Fig.~\ref{fig:gravwaves} displays the gravitational signal, represented by
the $l=2,m=2$ (the most dominant) component of $r M_{{\rm total}} \Psi_4$ (for a representative analysis of the late inspiral
GW from this binary see e.g.~\cite{Baiotti:2011am}).  This signal displays the standard ``chirping'' behavior 
in which the  amplitude and frequency grow as the orbit shrinks due to the radiation of angular momentum via
gravitational waves. The merger of these stars produces a hypermassive, differentially rotating
neutron star which emits gravitational waves as it radiates
excess energy and angular momentum before succumbing to collapse to a black hole.

As a result of the merger, as discussed in~\cite{Price:2006fi,2008PhRvL.100s1101A,Obergaulinger:2010gf}, 
magnetic fields can be amplified --via diverse mechanisms such as winding, Kevin-Helmholtz and MRI instabilities--
to values large enough that magnetic effects can indeed affect the {\em after-merger} dynamics, and thus the
corresponding gravitational wave signatures. To accurately resolve such
effects, resolutions at least an order of magnitude better are required. 
Furthermore, differences in the post-merger dynamics can arise from the choice of equation of state which also impacts
the waveform characteristics. We therefore  focus our analysis primarily up to the merger stage and discuss
briefly the early post-merger epoch, leaving for future work a closer examination 
of this late stage. We note however,
as discussed in~\cite{Lehner:2011aa,2013MNRAS.431.1853P}, that late-stage collapse can induce significant electromagnetic emission.

%---------------------------------------------

%---------------------------------------------

%---------------------------------------------

%---------------------------------------------

%%%%%%%%%%%%%%%%%%%%%%%%%%%%%%%%%%%%%%%%%%%%%%%%%%%%%%%%%%%%%%%
\subsection{Electromagnetic radiation and dissipation}
\label{subsec:EMResults}

In spite of the orbiting behavior being the same for all three cases considered,
the electromagnetic field dynamics and magnetosphere interactions 
naturally depend sensitively on the orientation of the magnetic dipole
moments of the stars.
Such interactions strongly affect the resulting topology of the global electromagnetic field and may induce
dissipation regions, reconnections of field lines, and a net Poynting flux,
as well as several other relevant features. At a rudimentary level, the
accelerated orbital motion
of the stars induces only a small degree of winding of its magnetic field
lines; thus the magnetospheres essentially co-rotate with the stars
and the magnetic field at their surfaces (and therefore, the magnetic 
dipole moments) remains almost constant until the merger. 
For the sake of clarity, we first discuss the main features of each case separately and
then compare and contrast particular aspects among the three cases. Henceforth we set $t=0$ as the
time at which the stars touch.

%%%%%%%%%%%%%%%%%%%%%%%%%%%%%%%%%%%%%%%%%%%%%%%%%%%%%%%%%%%%%%%

\subsubsection{U/D case}

\begin{figure}
%\begin{center}
%	\epsfig{file=./figures/Bflds_lines/bns03-Bfields_06.ps, width=0.45\columnwidth}
%	\epsfig{file=./figures/Bflds_lines/bns03-Bfields_12.ps, width=0.45\columnwidth}\\
%	\epsfig{file=./figures/Bflds_lines/bns03-Bfields_16.ps, width=0.45\columnwidth}
%	\epsfig{file=./figures/Bflds_lines/bns03-Bfields_20.ps, width=0.45\columnwidth}
	\epsfig{file=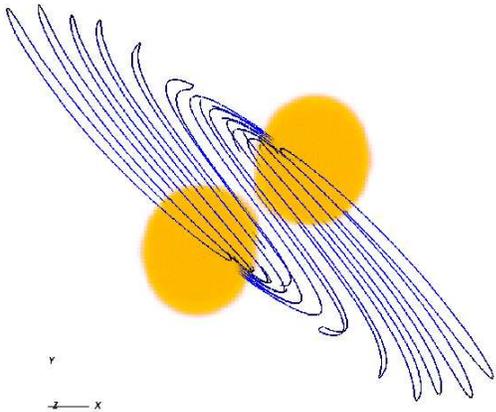, width=0.8\columnwidth}
	\caption{{\it U/D case}. Top-down view of certain magnetic field lines along with 
 the stellar density at $t=-1.7$ ms.
 Only field lines originating along a line connecting the stars slightly above the
 equatorial plane are shown for clarity.
              Similar plots for the other cases are shown in Fig.~\ref{fig:Bfield_UU} ({\it U/U})
                  and Fig.~\ref{fig:Bfield_Uu} ({\it U/u}).
 Note that reconnections near the leading edges of the stars (orbiting counter-clockwise in this view) have severed some lines
 which would otherwise connect the stars.
}
	\label{fig:Bfield_UD}
%\end{center}
\end{figure}

\begin{figure}
%\begin{center}
	\epsfig{file=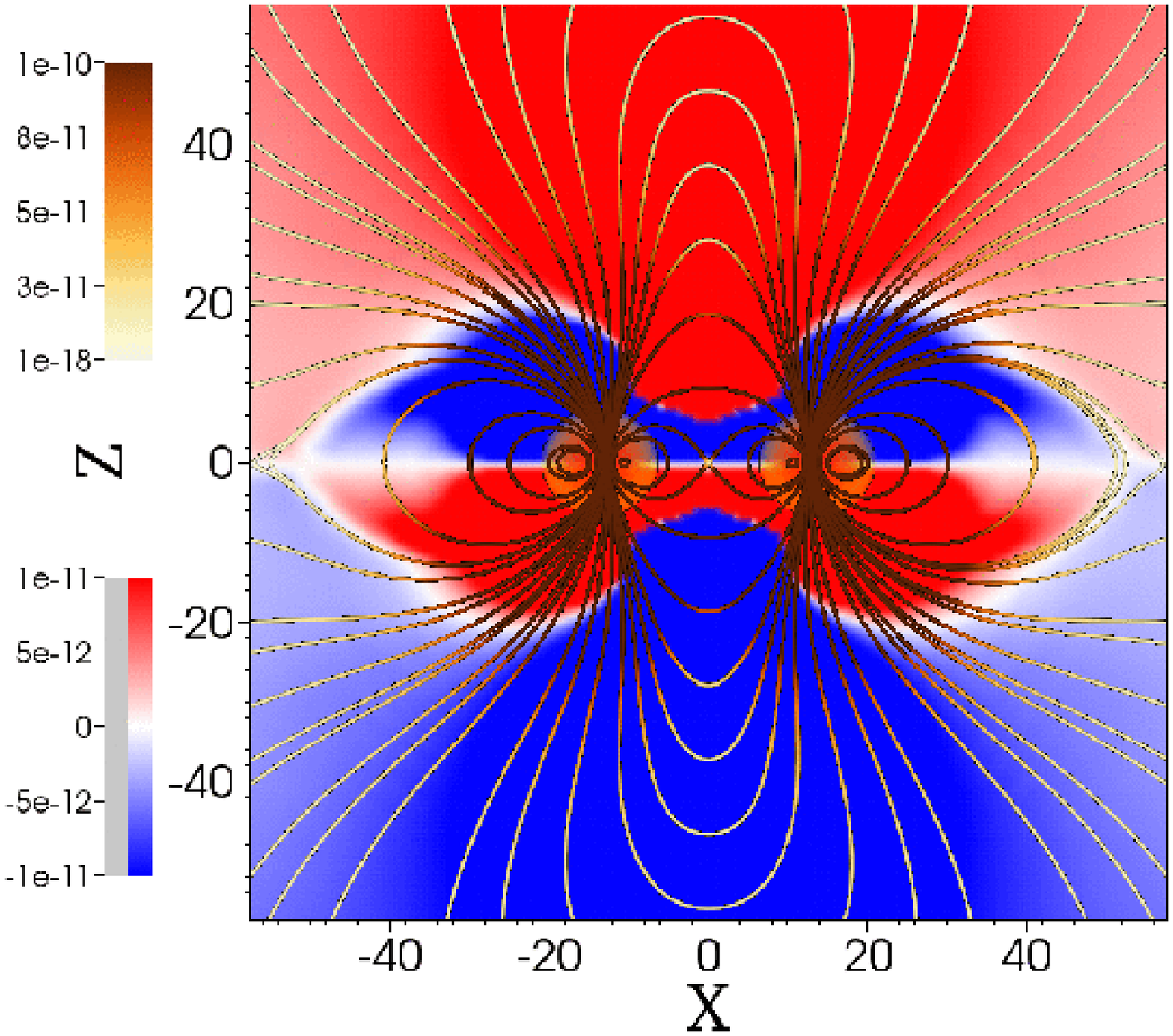, width=0.45\columnwidth}
	\epsfig{file=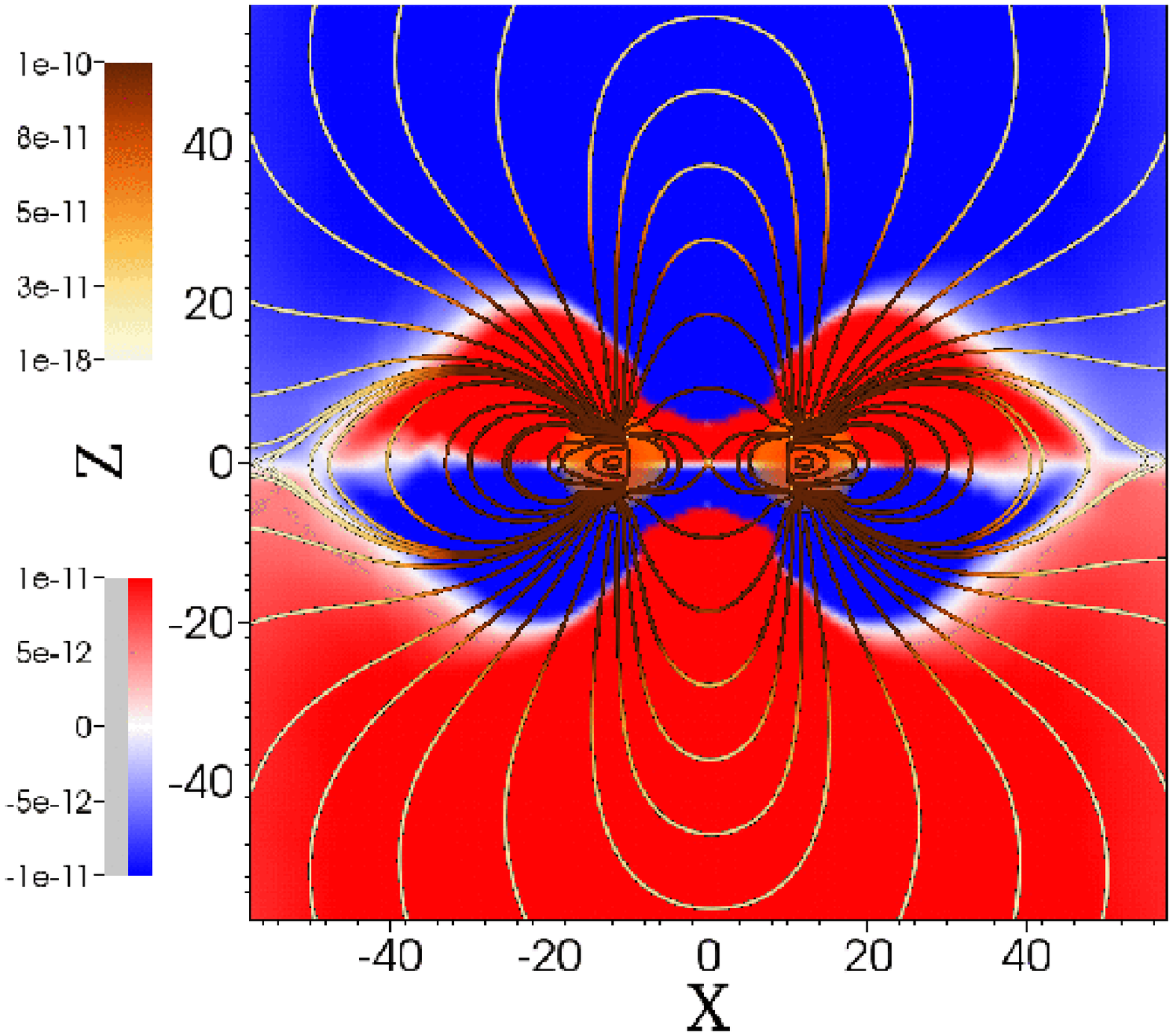, width=0.45\columnwidth}
	\epsfig{file=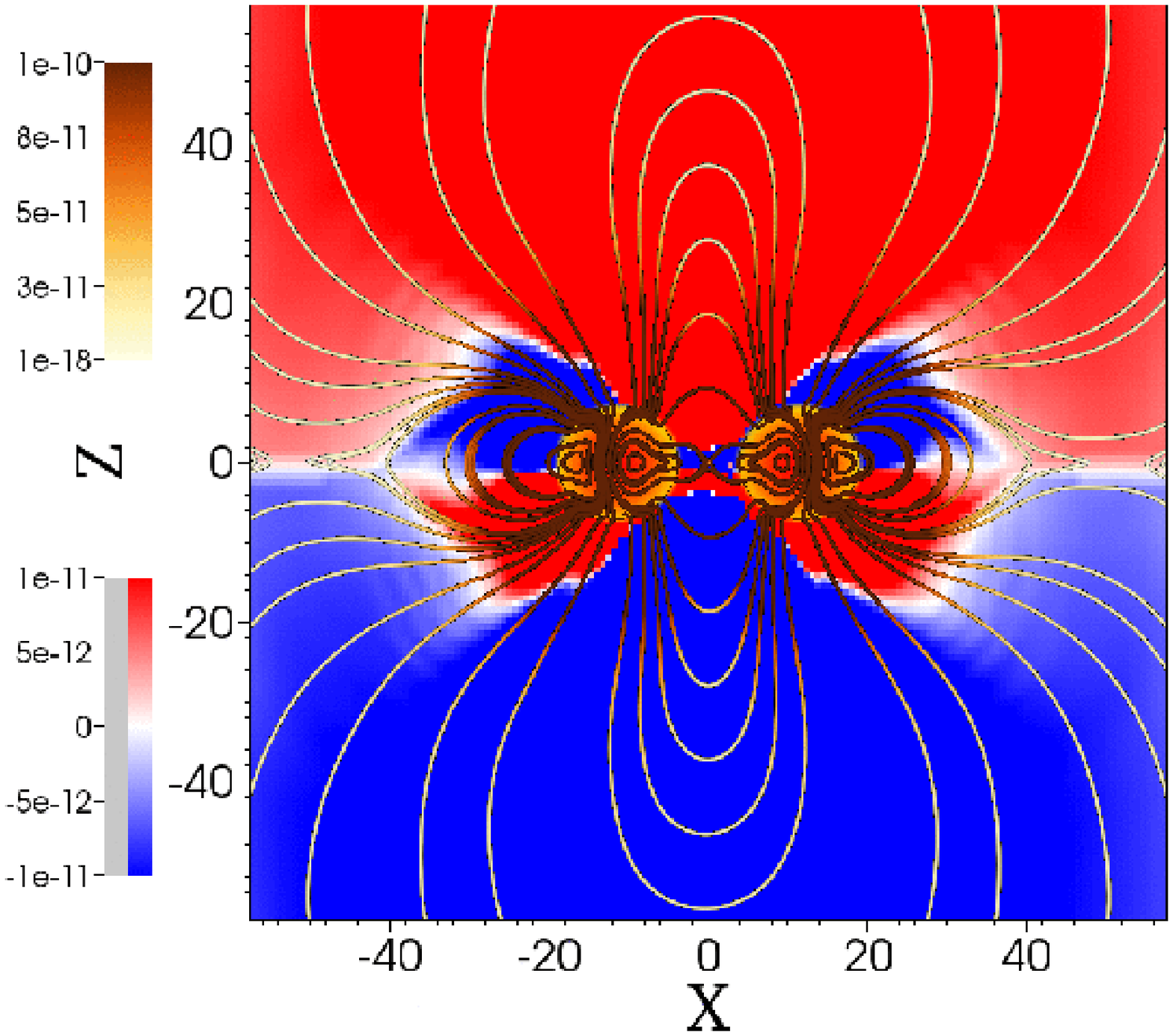, width=0.45\columnwidth}
	\epsfig{file=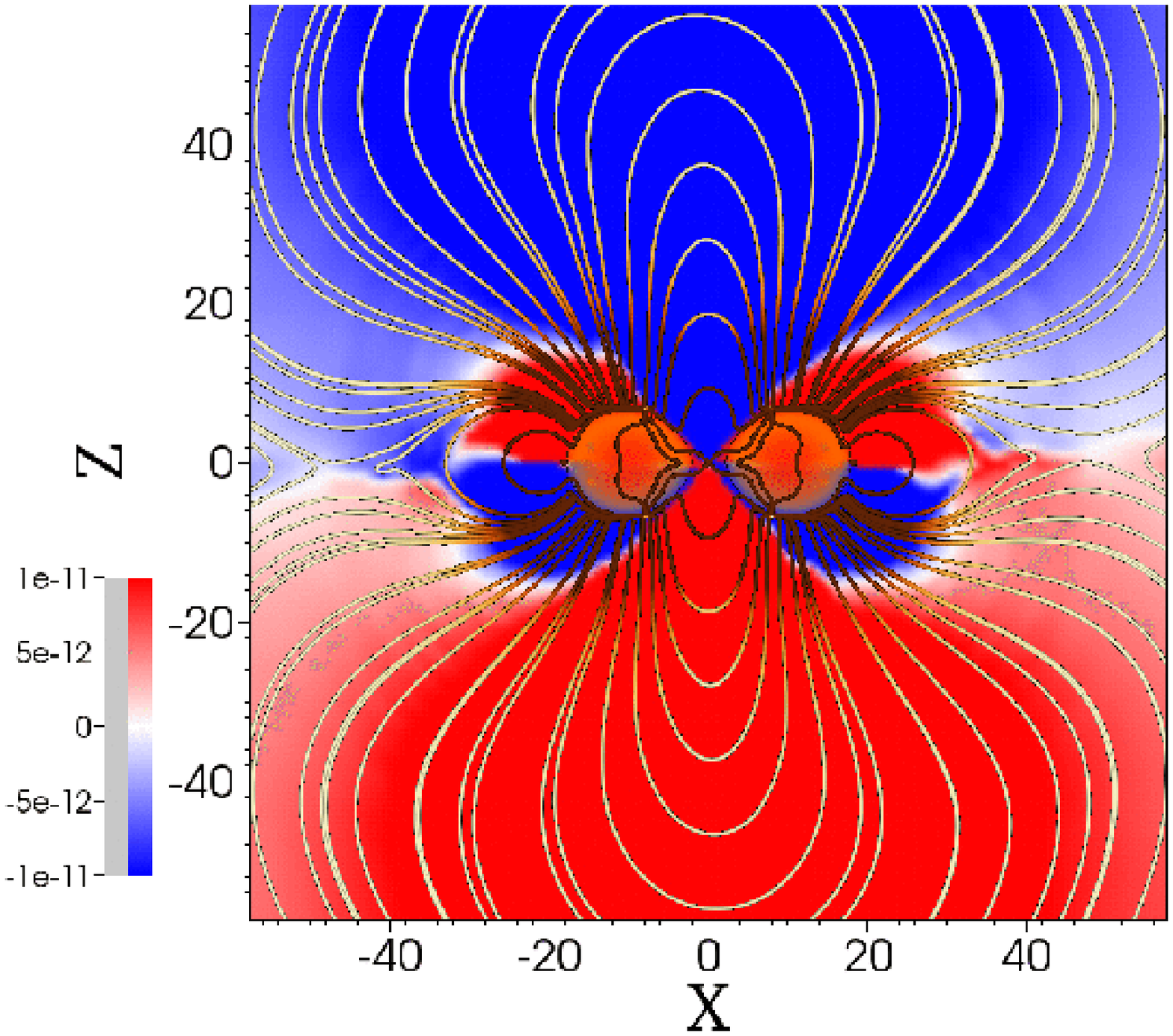, width=0.45\columnwidth}
	\caption{{\it U/D case}. Snapshots of the magnetic field configuration on the $y=0$ plane 
             at half orbital periods %OLD TIMES, keeping for ref: (times $t=1.8, 3.3, 4.7$ and $5.9$ ms)
             (times $t=-4.6, -3.2, -1.7$ and $-0.5$ ms).           
              Poloidal field lines are shown while the component perpendicular to the plane
              (related to the toroidal component) is shaded in color.
              Similar plots for the other cases are shown in Fig.~\ref{fig:Bfield_BcompUU} ({\it U/U})
                  and Fig.~\ref{fig:Bfield_BcompUu} ({\it U/u}).
              %Some field lines have reconnected and go from one star to the other.
              The field lines connecting the stars result from reconnection.
              Note that after each half-orbit, the direction of the perpendicular component
              switches direction due to the change in sign of the poloidal field.
}
	\label{fig:Bfield_BcompUD}
%\end{center}
\end{figure}

Fig.~\ref{fig:Bfield_UD} illustrates the behavior of the anti-aligned case~($U/D$)
in which both stars have equal magnitude magnetic moments but opposite directions--with
individual directions parallel and
antiparallel to the orbital angular momentum. The corotation of the magnetospheres
with the stars induces a shear layer in the midplane, separating two
regions filled with magnetically dominated plasma moving in opposite directions.
The poloidal component of the magnetic field from each star switches direction as one crosses
the midplane, allowing for reconnections that result in field lines connecting both stars.
The projection of these connecting field lines are quite apparent in Fig.~\ref{fig:Bfield_BcompUD}.

As the stars orbit, these field lines are severely stretched,
increasing their tension and developing a strong toroidal component.
Near the leading edge of each stellar surface,
these field lines undergo a twisting so extreme
that they are bent almost completely backwards, allowing them to reconnect and 
release some of the orbital energy stored by the twisted magnetic fields.
A careful examination of both Fig.~\ref{fig:Bfield_UD} and Fig.~\ref{fig:Bfield_BcompUD}
provides a view of S-shaped toroidal field lines connecting the stars.
The sense of the ``S'' changes as one crosses below the equatorial plane as shown in
Fig.~\ref{fig:Bfield_BcompUD}. As the stars orbit, our view of the S-shape changes
so that the colors switch with each half-orbit.

\begin{figure}
%\begin{center}
%%	\epsfig{file=./figures/Bflds_xi/bns03-Bfields_xi--12.ps, width=0.95\columnwidth}
%%        \epsfig{file=./figures/Bflds_xi/bns03-Bfields-xi_12_onB.ps, width=0.95\columnwidth}
%%        \epsfig{file=./figures/Bflds_xi/bns03-Bfields-xi_12_onW.ps, width=0.95\columnwidth}
        \epsfig{file=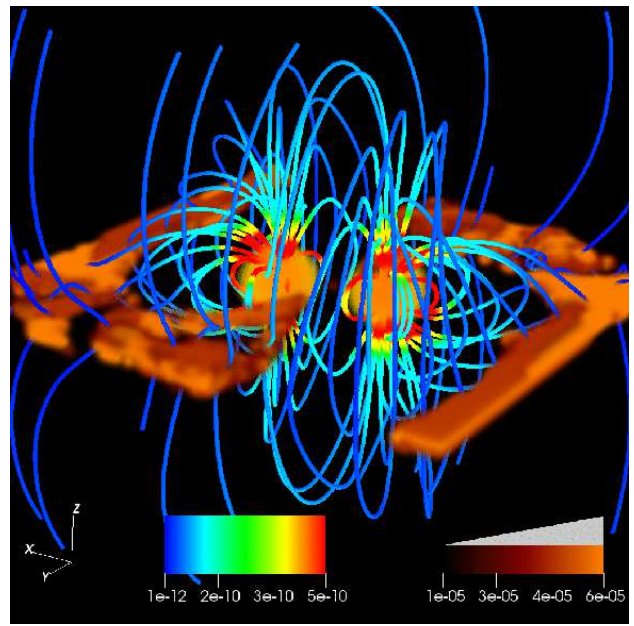, width=0.95\columnwidth}
	\caption{{\it U/D case}. Representative snapshot of the magnetic field configuration
        and induced current sheet  %OLD TIMES, keeping for ref: (at times $t=3.6$ and $4.7$ ms)
        at $t=-2.9$~ms.}
%        Bottom panels: Currents (drawn as arrows), magnetic field lines (solid lines),
%        and charge density (color coded) for the {\it U/D} case at %OLD TIMES, keeping for ref: $t=3.6$ and $4.7$ ms.
%        $t=-2.9$ and $-1.7$ ms.}
	\label{fig:currents-xi_UD}
%\end{center}
\end{figure}

The region near the stars, and especially near the orbital plane, is much more complicated. 
Reconnections at the midplane produce a current sheet
that propagates outwards, forming a spiral pattern. 
The current sheet structure is shown in Fig.~\ref{fig:currents-xi_UD},
and it rotates with the periodicity of the orbital motion, 
Assuming that the current sheet would produce electromagnetic radiation, then this
structure may effectively provide a ``spacetime tracer'' of the orbital motion.

The dynamics also impact the distribution of charges and currents as
illustrated in Fig.~\ref{fig:currents-3cases}. The top panel of this figure
shows the currents of the $U/D$ case with arrows. 
A circuit is established, with current flowing from the star on the left
to the one on the right well above and below the equator, with a returning
current closer to the equator.
These currents become stronger as the stars get closer.

%---------------------------------------------
%\begin{widetext}
%\end{widetext}
%---------------------------------------------

%---------------------------------------------
%\begin{widetext}
%\end{widetext}
%---------------------------------------------

%---------------------------------------------
%\begin{widetext}
%\end{widetext}
%---------------------------------------------

%%%%%%%%%%%%%%%%%%%%%%%%%%%%%%%%%%%%%%%%%%%%%%%%%%%%%%%%%%%%%%%

\subsubsection{U/U case}
\begin{figure}
%\begin{center}
%	\epsfig{file=./figures/Bflds_lines/bns01-Bfields_06.ps, width=0.45\columnwidth}
%	\epsfig{file=./figures/Bflds_lines/bns01-Bfields_12.ps, width=0.45\columnwidth}\\
%	\epsfig{file=./figures/Bflds_lines/bns01-Bfields_16.ps, width=0.45\columnwidth}
%	\epsfig{file=./figures/Bflds_lines/bns01-Bfields_20.ps, width=0.45\columnwidth}
	\epsfig{file=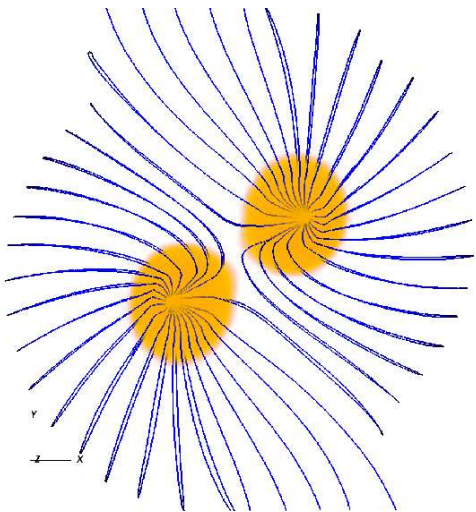, width=0.8\columnwidth}
	\caption{{\it U/U case}. Top-down view of certain magnetic field lines along with 
 the stellar density at $t=-1.7$ ms.
              Similar plots for the other cases are shown in Fig.~\ref{fig:Bfield_UD} ({\it U/D})
                  and Fig.~\ref{fig:Bfield_Uu} ({\it U/u}).
                                             The repulsion of roughly aligned
         field lines is clearly visible at the midplane between the stars.}
	\label{fig:Bfield_UU}
%\end{center}
\end{figure}

We turn now to general features displayed by the aligned ($U/U$) case in which 
both stars have identical magnetic moments, parallel to the orbital angular momentum.
A sense of the magnetic field structure for this case is given by
Fig.~\ref{fig:Bfield_UU}. Similar to
the $U/D$ case, a shear layer in the midplane between the stars arises
due to the oppositely directed velocities of the magnetospheres 
However, unlike the $U/D$ case, the magnetic field on both sides of the shear layer generally
point in the same direction and therefore do not reconnect.
 
\begin{figure}
%\begin{center}
	\epsfig{file=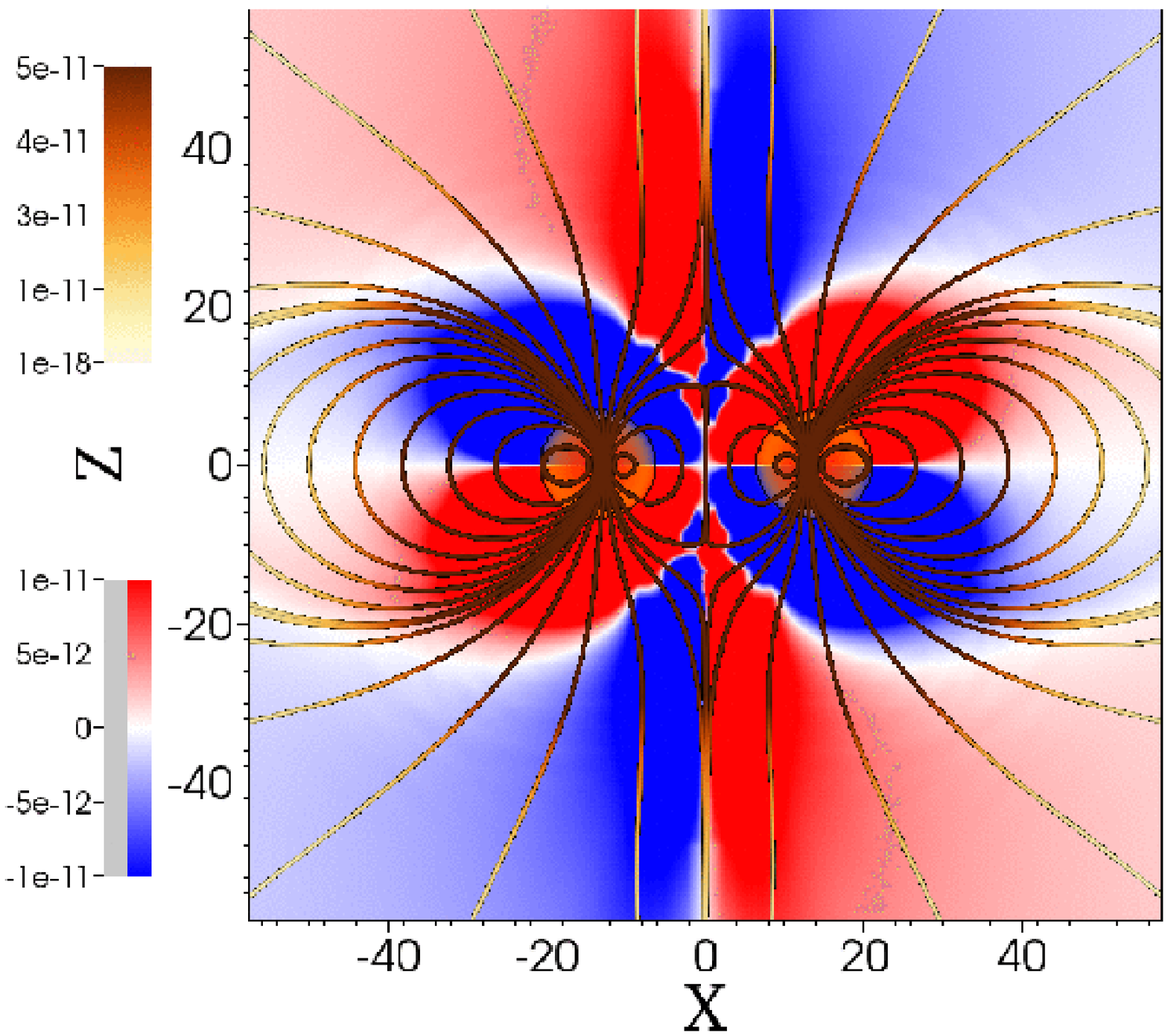, width=0.45\columnwidth}
	\epsfig{file=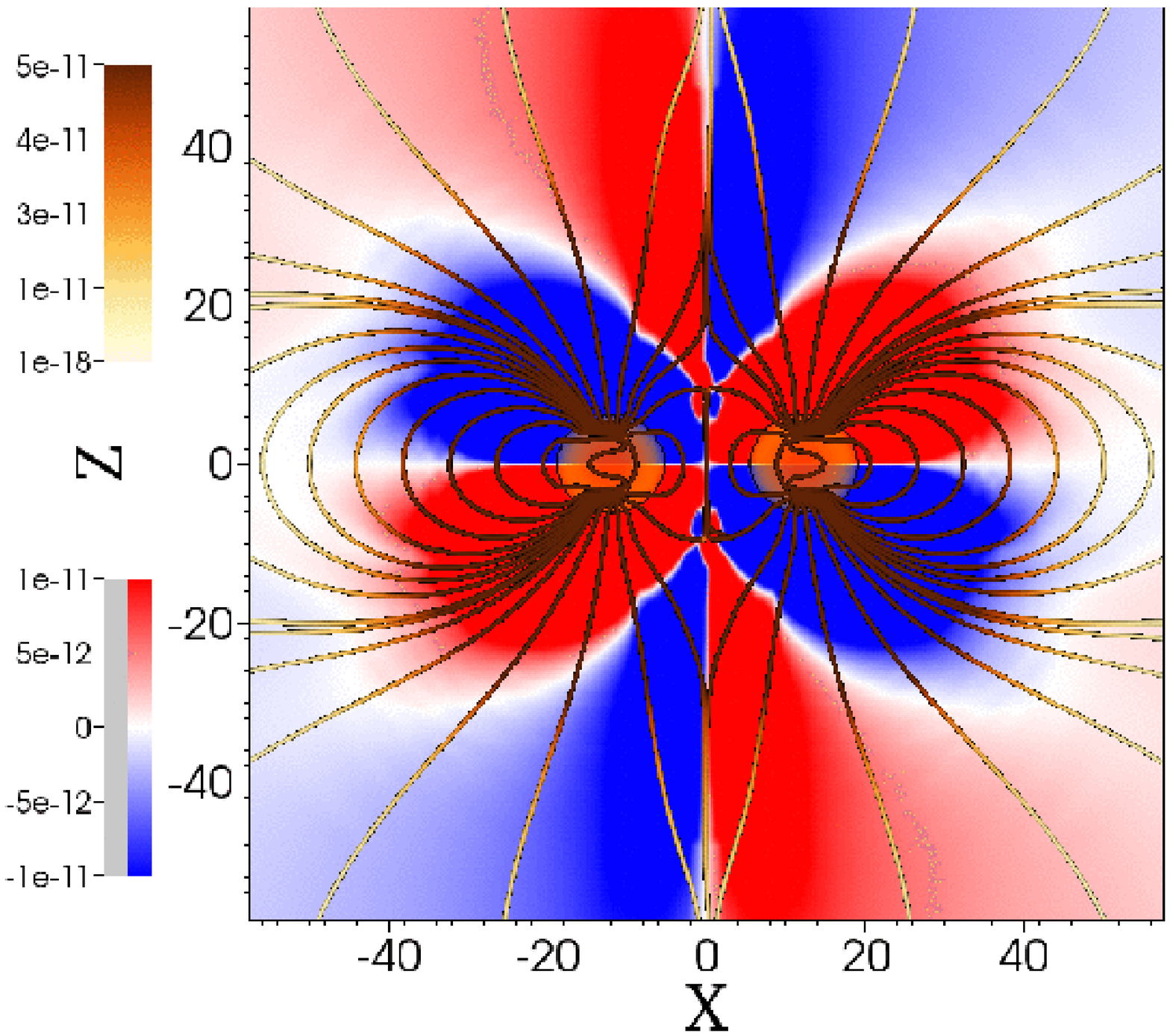, width=0.45\columnwidth}
	\epsfig{file=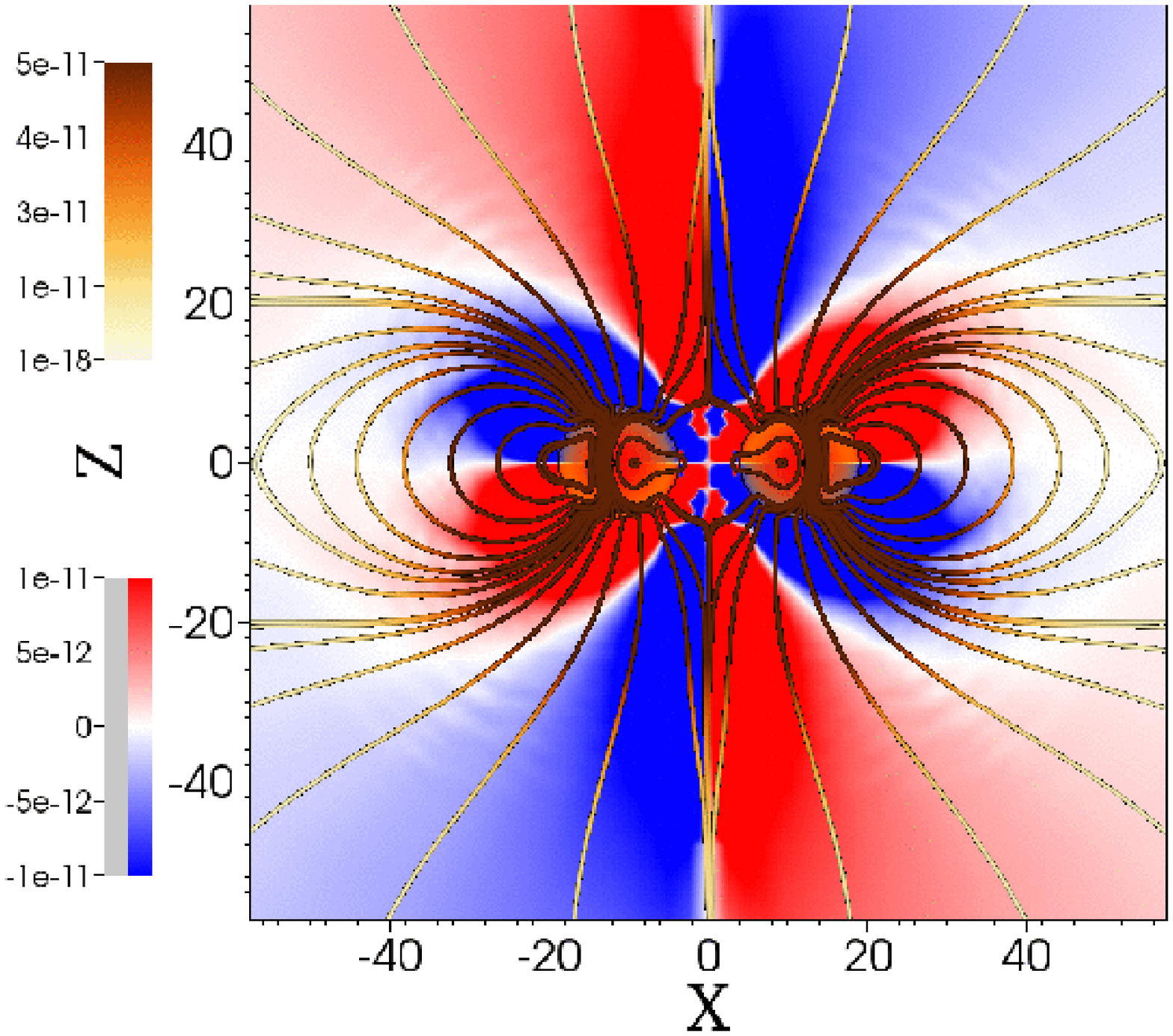, width=0.45\columnwidth}
	\epsfig{file=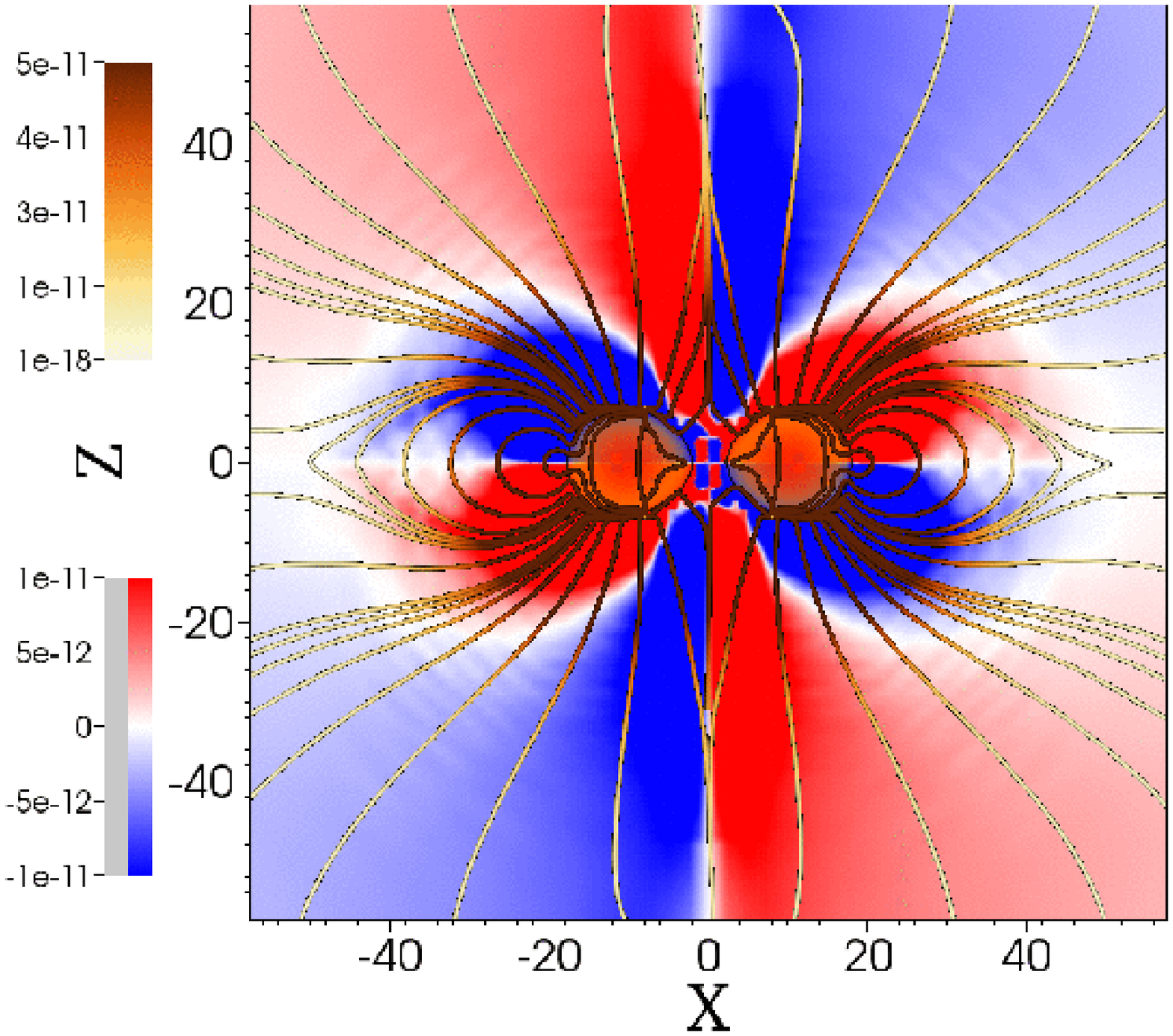, width=0.45\columnwidth}
	\caption{{\it U/U case}. Snapshots of the magnetic field configuration on the $y=0$ plane 
             at half orbital periods  %OLD TIMES, keeping for ref: (times $t=1.8, 3.3, 4.7$ and $5.9$ ms) for the {\it U/U} case.
             (times $t=-4.6, -3.2, -1.7$ and $-0.5$ ms).
              Similar plots for the other cases are shown in Fig.~\ref{fig:Bfield_BcompUD} ({\em U/D})
                  and Fig.~\ref{fig:Bfield_BcompUu} ({\em U/u}).
              Note that field lines near the midplane repel each other as is more evident in
              Fig.~\ref{fig:Bfield_UU}.
              Far from the binary at large radii, the magnetic field structure resembles that of
              an aligned rotator.
              Indeed the direction of the magnetic field changes across the
              equatorial plane and a Y-point arises as the orbit tightens.   
}
	\label{fig:Bfield_BcompUU}
%\end{center}
\end{figure}

Again, rotation induces a toroidal magnetic field although the 
topology is clearly different than the one observed in the $U/D$ case. Here,
the deflection of the field lines from each star at the midplane produces
a strong, antisymmetric toroidal component in the central region of the midplane.
Far from the stars, the structure is reminiscent of the one obtained in the aligned (dipole)
rotator (see e.g.~\cite{Spitkovsky:2006np,2012arXiv1211.2803T,2013MNRAS.431.1853P}).
One aspect of this rotator structure is the appearance of a Y-point in the poloidal field along
the equatorial plane roughly at large radius (see the points near $(\pm 45,0))$ in the last panel of Fig.~\ref{fig:Bfield_BcompUU}). This point separates 
closed from open field lines and occurs at the {\em light cylinder}, the radius
$R_{\rm LC}=c/\Omega$ at which the velocity of the co-rotating magnetosphere reaches light speed. Another aspect similar to the rotator is oppositely directed
toroidal field as one cross the equatorial plane.
This structure is natural, as there is a net effective dipole to leading order in the system. 
However in this case the symmetry of
the system implies an (approximate) periodicity in the solution given
by half the orbital period, which is more evident in Fig.~\ref{fig:Bfield_BcompUU}. 

\begin{figure}
%\begin{center}
%	\epsfig{file=./figures/Bflds_xi/bns01-Bfields_xi--06.ps, width=0.5\columnwidth}
%%	\epsfig{file=./figures/Bflds_xi/bns01-Bfields_xi--12.ps, width=0.95\columnwidth}
%%        \epsfig{file=./figures/Bflds_xi/bns01-Bfields-xi_12_onB.ps, width=0.95\columnwidth}
%%        \epsfig{file=./figures/Bflds_xi/bns01-Bfields-xi_12_onW.ps, width=0.95\columnwidth}
        \epsfig{file=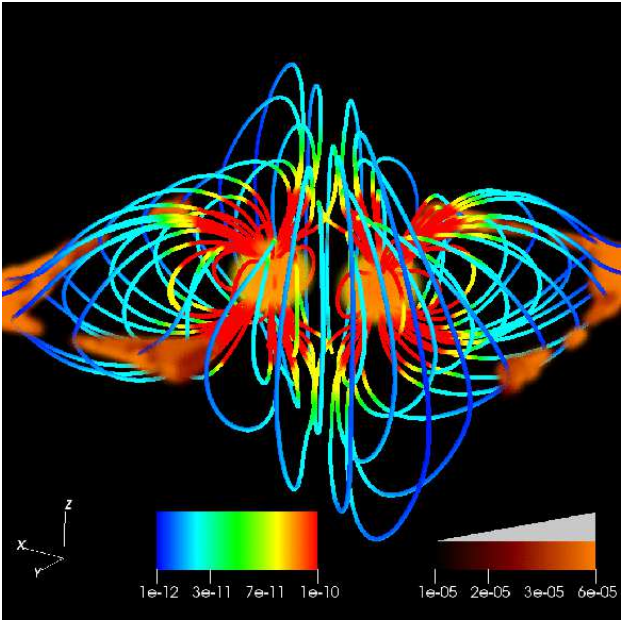, width=0.95\columnwidth}
	\caption{{\it U/U case}. Snapshot of the magnetic field configuration
         and induced current sheet  %OLD TIMES, keeping for ref: (at times $t=3.6$ and $4.7$ ms)
         at $t=-2.9$ ms.}
%	 Bottom panels: Currents (drawn as arrows) and charge density (color
%         coded) for the {\it U/U} case at %OLD TIMES, keeping for ref: $t=3.6$ and $4.7$ ms.
%         $t=-2.9$ and $-1.7$ ms.}
	\label{fig:currents-xi_UU}
%\end{center}
\end{figure}

Furthermore, a current sheet is induced on the equatorial plane as
shown in Fig.~\ref{fig:currents-xi_UU}.
The current sheet first arises at the light cylinder which shrinks
as the orbit tightens, resembling that of an aligned rotator. 
Once again the current sheet reflects the dynamics of the binary and hence may be a tracer
of the spacetime. In particular, the current sheet is
not homogeneous along the azimuthal direction, presenting a spiral pattern. 
The induced current distribution and charge density are displayed in 
the middle panel of Fig.~\ref{fig:currents-3cases}, revealing a
strong current along the $z$-axis towards the center surrounded by a sheath of outwardly
directed current.

%---------------------------------------------
%\begin{widetext}
%\end{widetext}
%---------------------------------------------

%---------------------------------------------
%\begin{widetext}
%\end{widetext}
%---------------------------------------------

%---------------------------------------------
%\begin{widetext}
%\end{widetext}
%---------------------------------------------

%%%%%%%%%%%%%%%%%%%%%%%%%%%%%%%%%%%%%%%%%%%%%%%%%%%%%%%%%%%%%%%

\subsubsection{U/u case}

\begin{figure}
%\begin{center}
	\epsfig{file=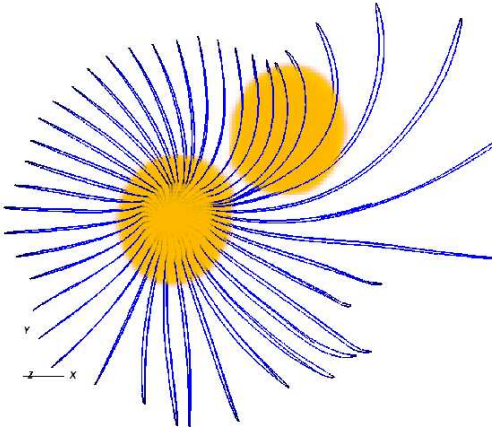, width=0.8\columnwidth}
	\caption{{\it U/u case}. Top-down view of certain magnetic field lines along with 
 the stellar density at $t=-1.7$ ms.
              Similar plots for the other cases are shown in Fig.~\ref{fig:Bfield_UD} ({\it U/D})
                  and Fig.~\ref{fig:Bfield_UU} ({\it U/U}).
	%\caption{{\it U/u case}. Top view of the star's density and the magnetic field
         %configuration at $t=-1.7$ ms. 
         Note that field lines emanating from the
         strongly magnetized star bend both around the weaker star 
         and the region just in front of it (the stars are orbiting
         counterclockwise). Similarly, the
         lines behind the weaker star are  distorted by a trailing current sheet.}
	\label{fig:Bfield_Uu}
%\end{center}
\end{figure}

The last of the three cases, $U/u$, contains one star significantly less
magnetized (just $1\%$) than the other, although both moments are initially aligned.
Among these cases, this $U/u$ case most resembles the models of~\cite{2001MNRAS.322..695H,Piro:2012rq,Lai:2012qe}
which study a binary with just one star initially magnetized.
These models invoke the unipolar inductor as discussed earlier in Section~\ref{sec:unipolar}.
As mentioned there, when the stars are well separated, the field produced by the
weaker star shields the star from the more magnetized field within some effective
radius.
Fig.~\ref{fig:Bfield_Uu} illustrates that the magnetic field
from the strongly magnetized star eventually dominates that of the companion.
As a consequence, the global electromagnetic field for the system is roughly described by
an inspiraling, magnetic dipole perturbed by induction effects on the weaker star.

\begin{figure}
%\begin{center}
	\epsfig{file=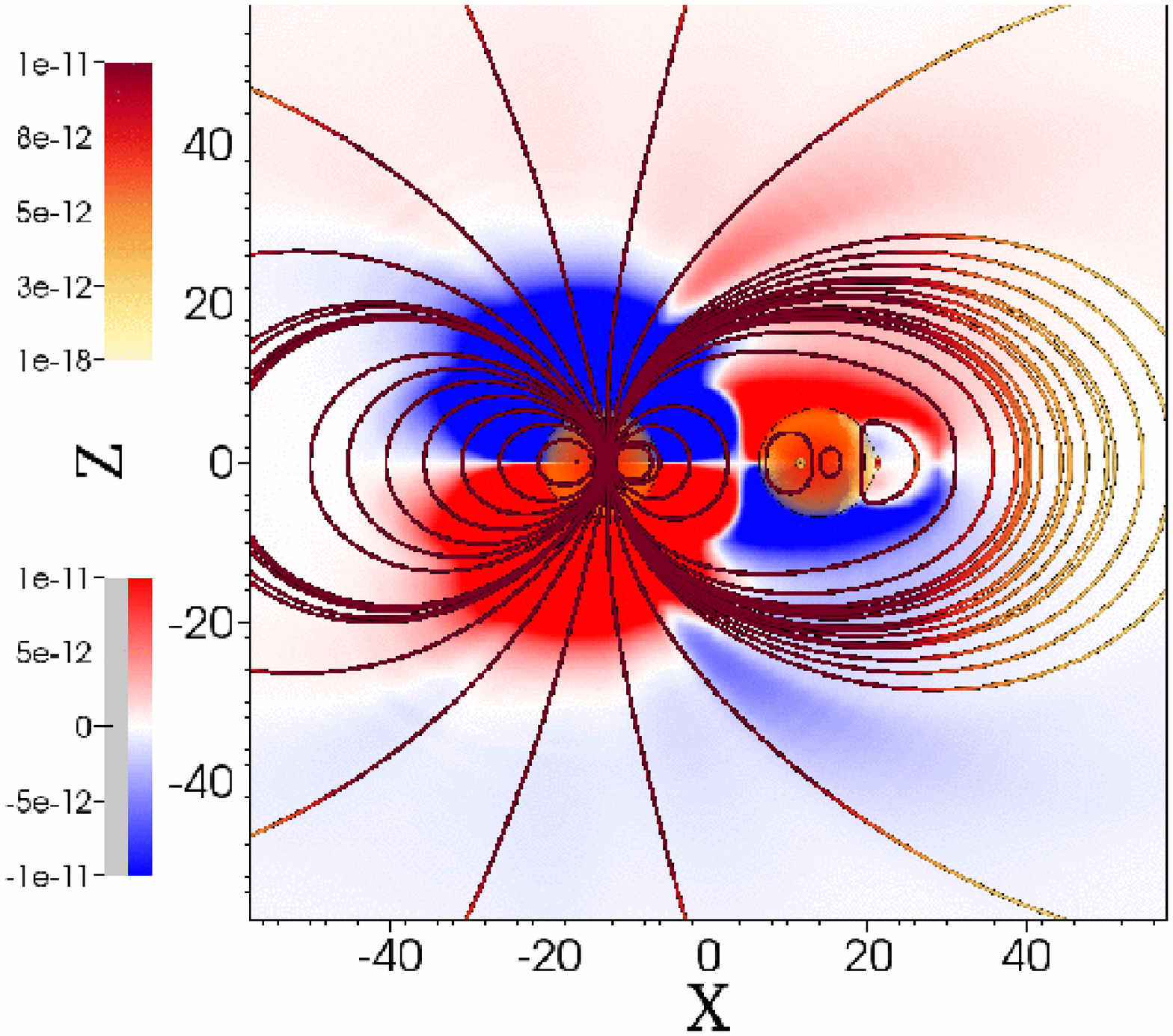, width=0.45\columnwidth}
	\epsfig{file=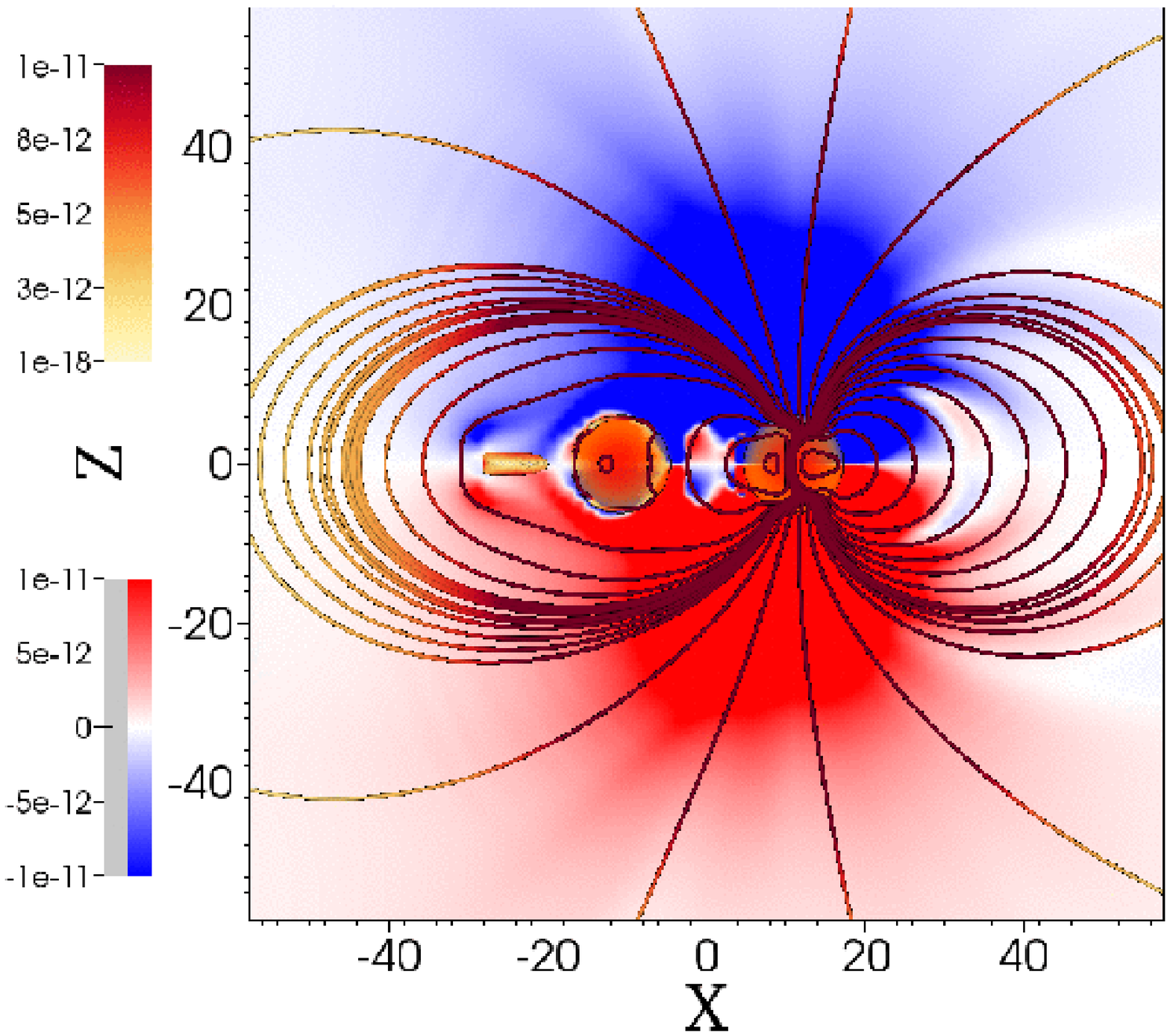, width=0.45\columnwidth}
	\epsfig{file=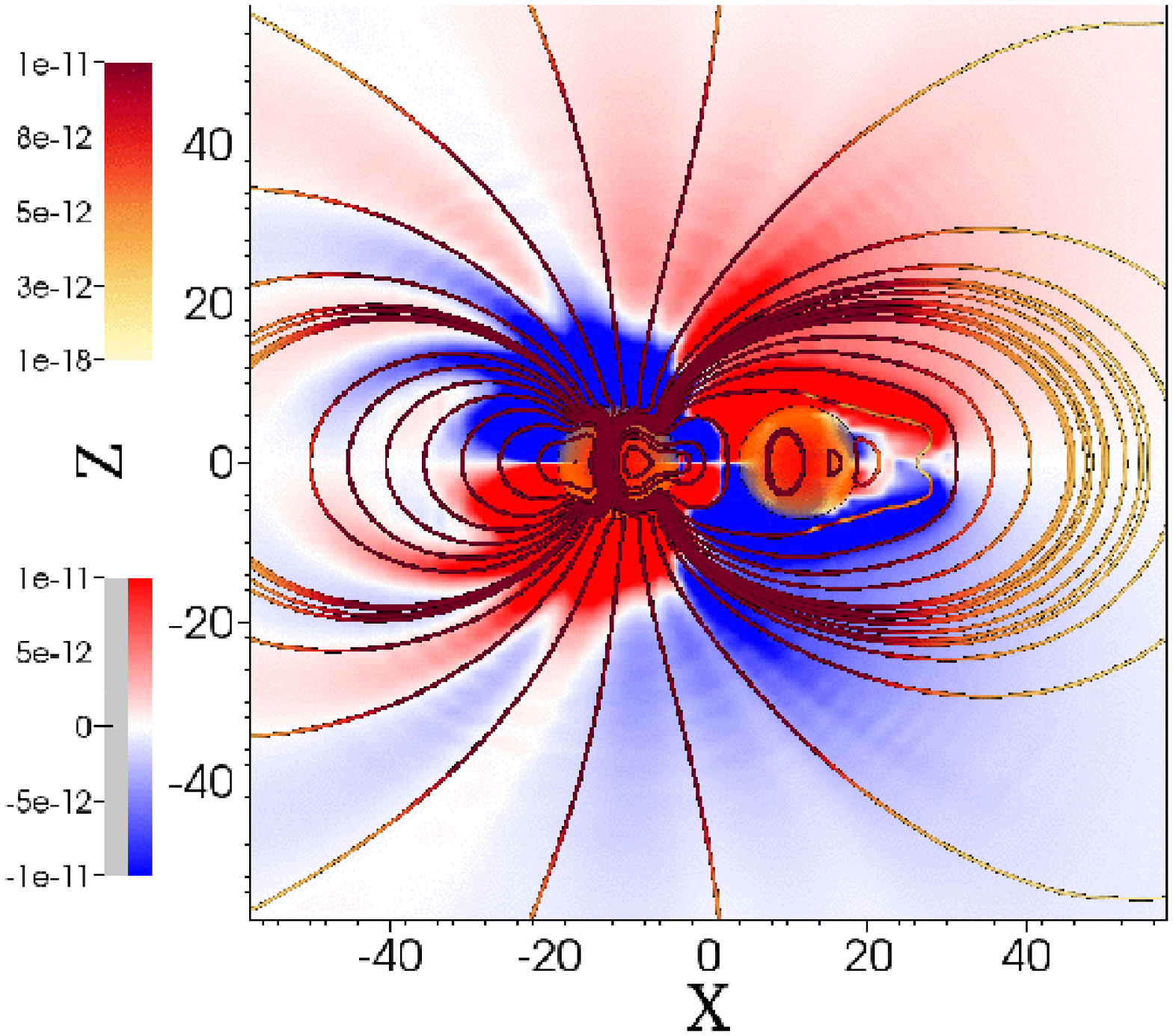, width=0.45\columnwidth}
	\epsfig{file=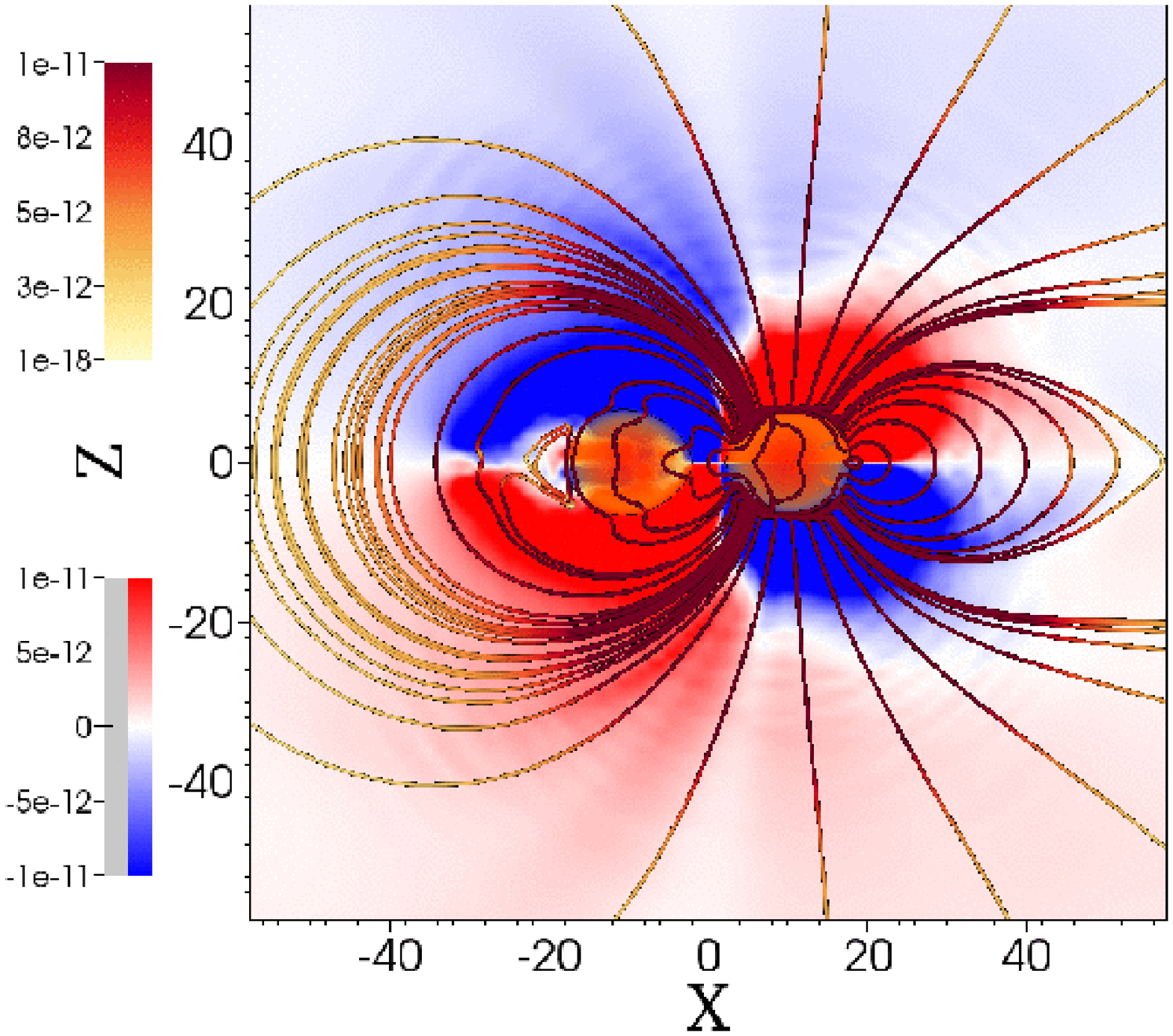, width=0.45\columnwidth}
	\caption{{\it U/u case}. Snapshots of the magnetic field configuration on the $y=0$ plane at
             half orbital periods  %OLD TIMES, keeping for ref: (times $t=1.8, 3.3, 4.7$ and $5.9$ ms) for the {\it U/u} case.
              (times $t=-4.6, -2.9, -1.7$ and $-0.5$ ms).
              Similar plots for the other cases are shown in Fig.~\ref{fig:Bfield_BcompUD} ({\it U/D})
                  and Fig.~\ref{fig:Bfield_BcompUU} ({\it U/U}).
Notice that the magnetic field structure is mainly described by an
orbiting dipole perturbed by the interaction with the
weakly magnetized companion and its trailing current sheet.
}
	\label{fig:Bfield_BcompUu}
%\end{center}
\end{figure}

The magnetic field for the $U/u$ case is also shown in Fig.~\ref{fig:Bfield_BcompUu}.
Note that in the first panel, one can see that the weaker field shields the
less magnetized star (on right) from that of the dominant star.

\begin{figure}
%\begin{center}
%	\epsfig{file=./figures/Bflds_xi/bns02-Bfields_xi--06.ps, width=0.5\columnwidth}
%	\epsfig{file=./figures/Bflds_xi/bns02-Bfields-xi_12.ps, width=0.45\columnwidth}
%%	\epsfig{file=./figures/Bflds_xi/bns02-Bfields-xi_16.ps, width=0.95\columnwidth}
%%        \epsfig{file=./figures/Bflds_xi/bns02-Bfields-xi_16_onB.ps, width=0.95\columnwidth}
%%        \epsfig{file=./figures/Bflds_xi/bns02-Bfields_xi-16_onW.ps, width=0.95\columnwidth}
        \epsfig{file=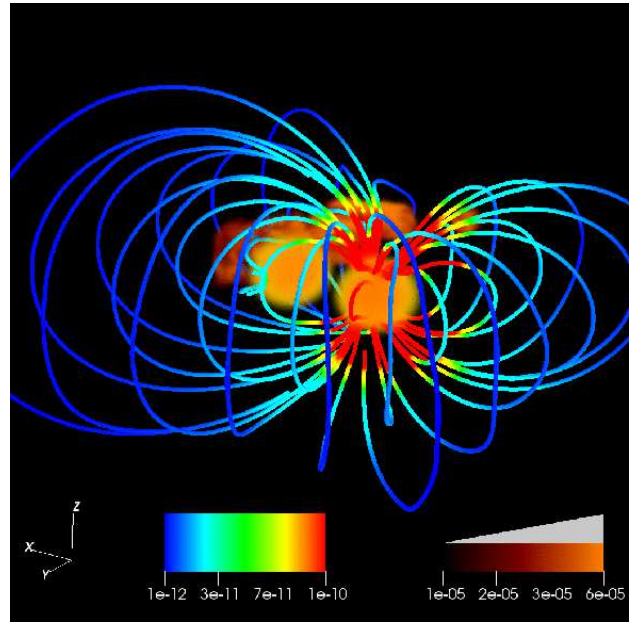, width=0.95\columnwidth}
	\caption{{\it U/u case}. Snapshot of the magnetic field configuration 
         and current sheet at % $t=-2.9$
         $t=-1.7$ ms.}
%	 Bottom panels: Currents (drawn as arrows) and charge density (color
%         coded) for the {\it U/u} case at %OLD TIMES, keepig for ref: $t=3.6$ and $4.7$ ms.
%         $t=-2.9$ and $-1.7$ ms.}
	\label{fig:currents-xi_Uu}
%\end{center}
\end{figure}

An interesting effect occurs as magnetic field lines originating from the strongly magnetized star 
slide off the companion's surface and then reconnect. This reconnection
produces a trailing region of dissipation, quite visible in  
Fig.~\ref{fig:currents-xi_Uu}. The extent of this dissipative tail gradually grows as the
stars orbit, populating a current sheet.

\begin{figure}
%\begin{center}
	\epsfig{file=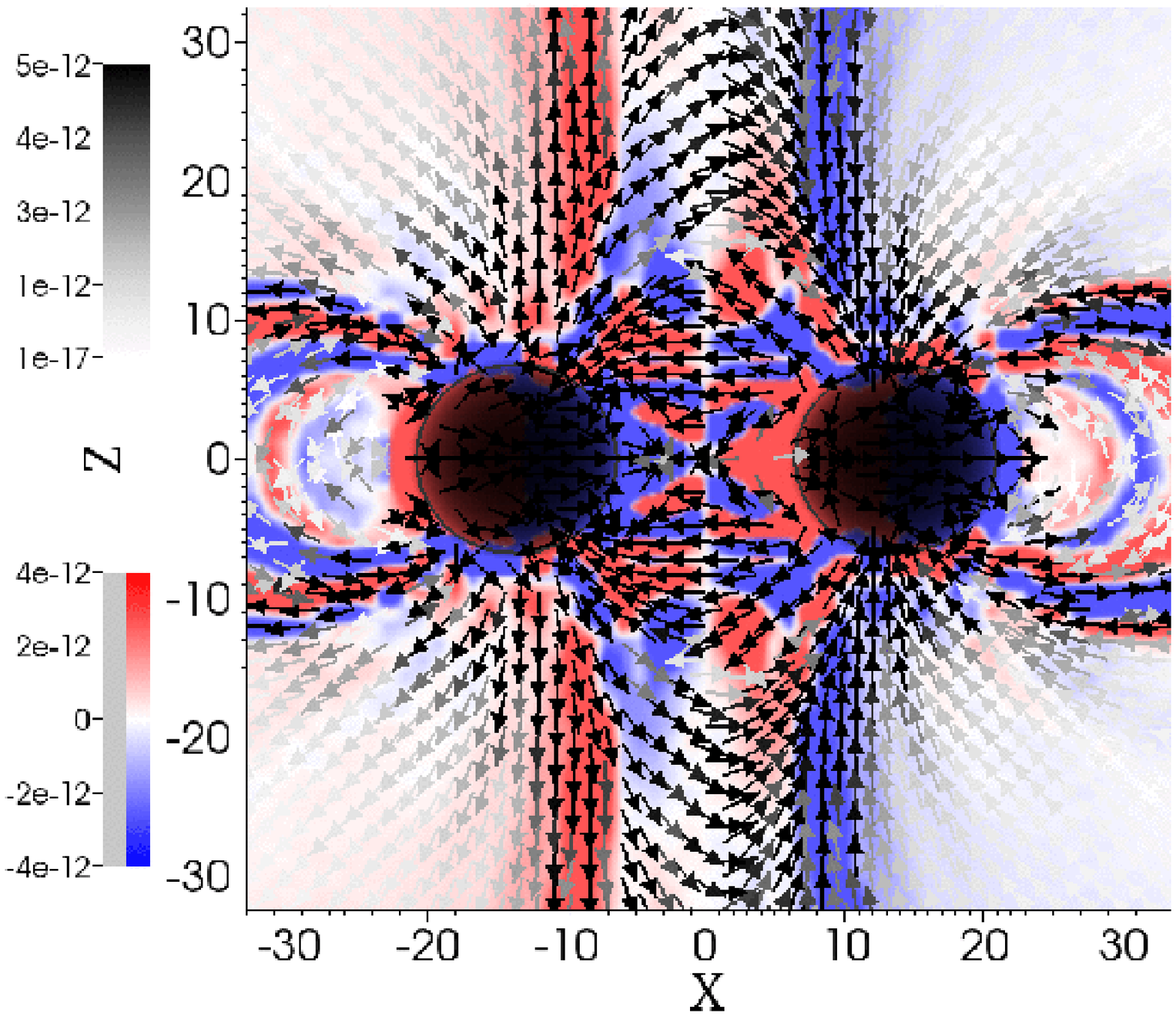, width=0.68\columnwidth}
	\epsfig{file=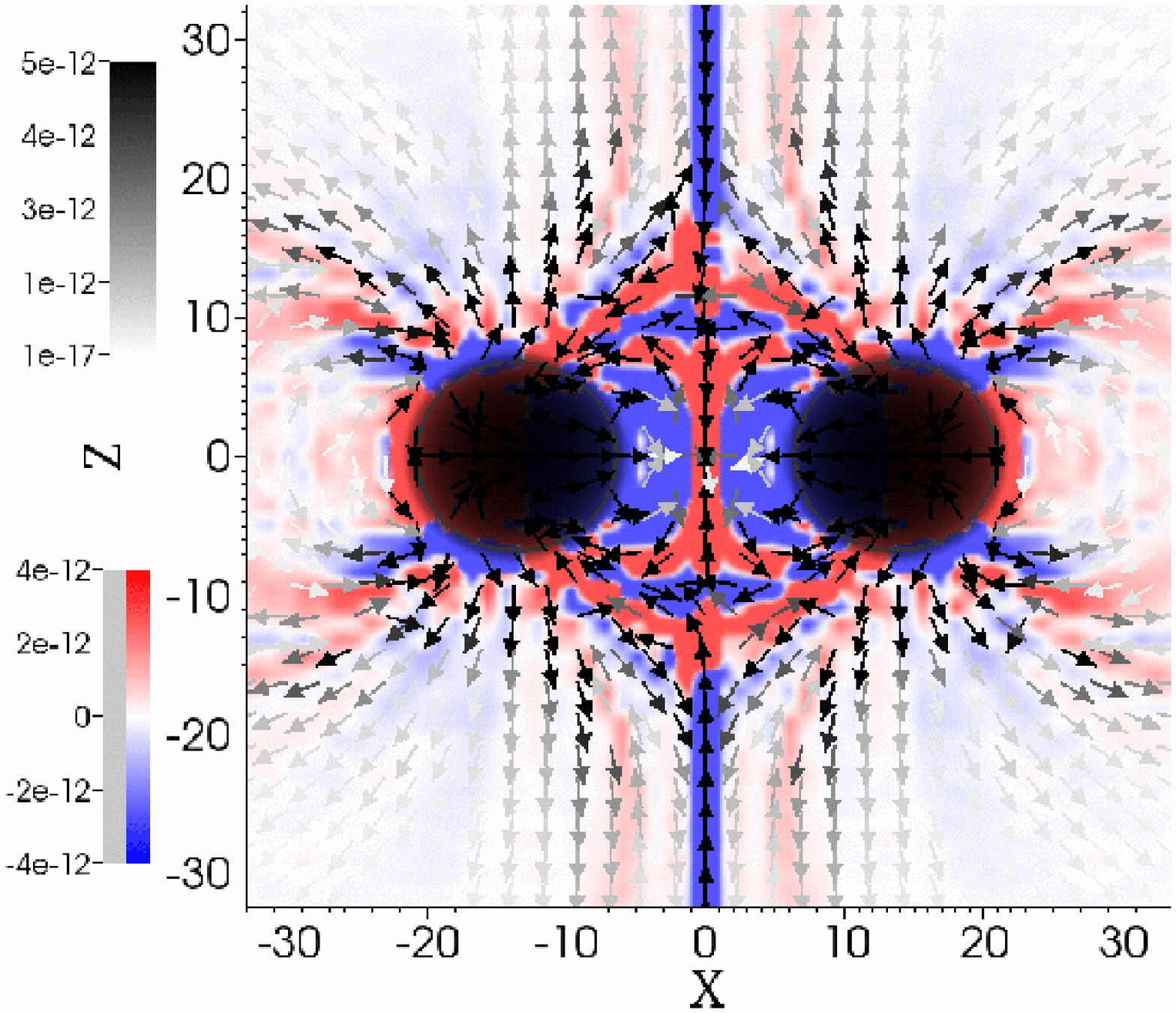, width=0.68\columnwidth}
        \epsfig{file=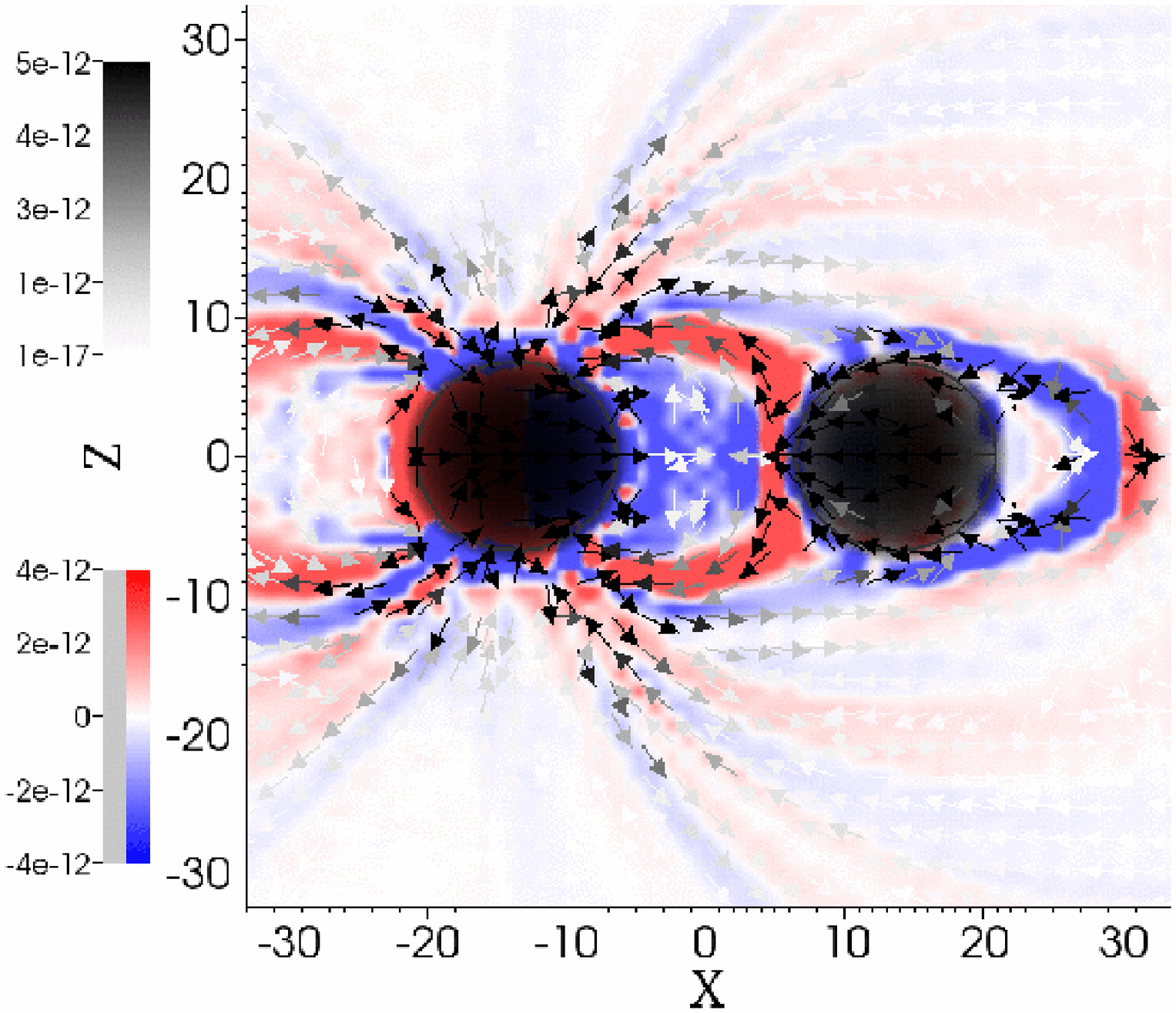, width=0.68\columnwidth}
	\caption{Currents (drawn as arrows) and charge density (color coded) for the 
{\it U/D} (first panel), {\it U/U} (second panel) and {\it U/u} (third panel) case at
$t = -4.6$ ms.	%$t = -0.5$ ms. %OLD TIME: $t = 5.9$ ms.
In all cases an effective circuit arises; however the circuits extend significantly
in both vertical directions for the first two cases which contrasts with the more
localized circuit in the last case. The bottom panel resembles the diagram in
Fig.~1 of~\cite{Lai:2012qe}.
}
	\label{fig:currents-3cases}
%\end{center}
\end{figure}

The induced current and charge density reveal a structure consistent with
the unipolar induction model, as can be seen in the bottom panel of
Fig.~\ref{fig:currents-3cases}.
A closed circuit between the stars is shown, with current flowing
along the magnetic field lines from the strongly magnetized star to the rear
of the weakly magnetized one. This current makes its return along the stellar surface
until following the returning field lines back to the strongly magnetized star
to complete the circuit (see for instance the diagram in Fig.~1 of~\cite{Lai:2012qe}).

%---------------------------------------------
%\begin{widetext}
%\end{widetext}
%---------------------------------------------

%---------------------------------------------
%\begin{widetext}
%\end{widetext}
%---------------------------------------------

%---------------------------------------------
%\begin{widetext}
%\end{widetext}
%---------------------------------------------

%%%%%%%%%%%%%%%%%%%%%%%%%%%%%%%%%%%%%%%%%%%%%%%%%%%%%

%%%%%%%%%%%%%%%%%%%%%%%%%%%%%%%%%%%%%%%%%%
\subsection*{Poynting Flux and Energy Dissipation}

\begin{figure}
%\begin{center}
%	\epsfig{file=./figures/sqphi2/bns03_UD-sqphi2_3d-20.ps, width=0.48\columnwidth}
	\epsfig{file=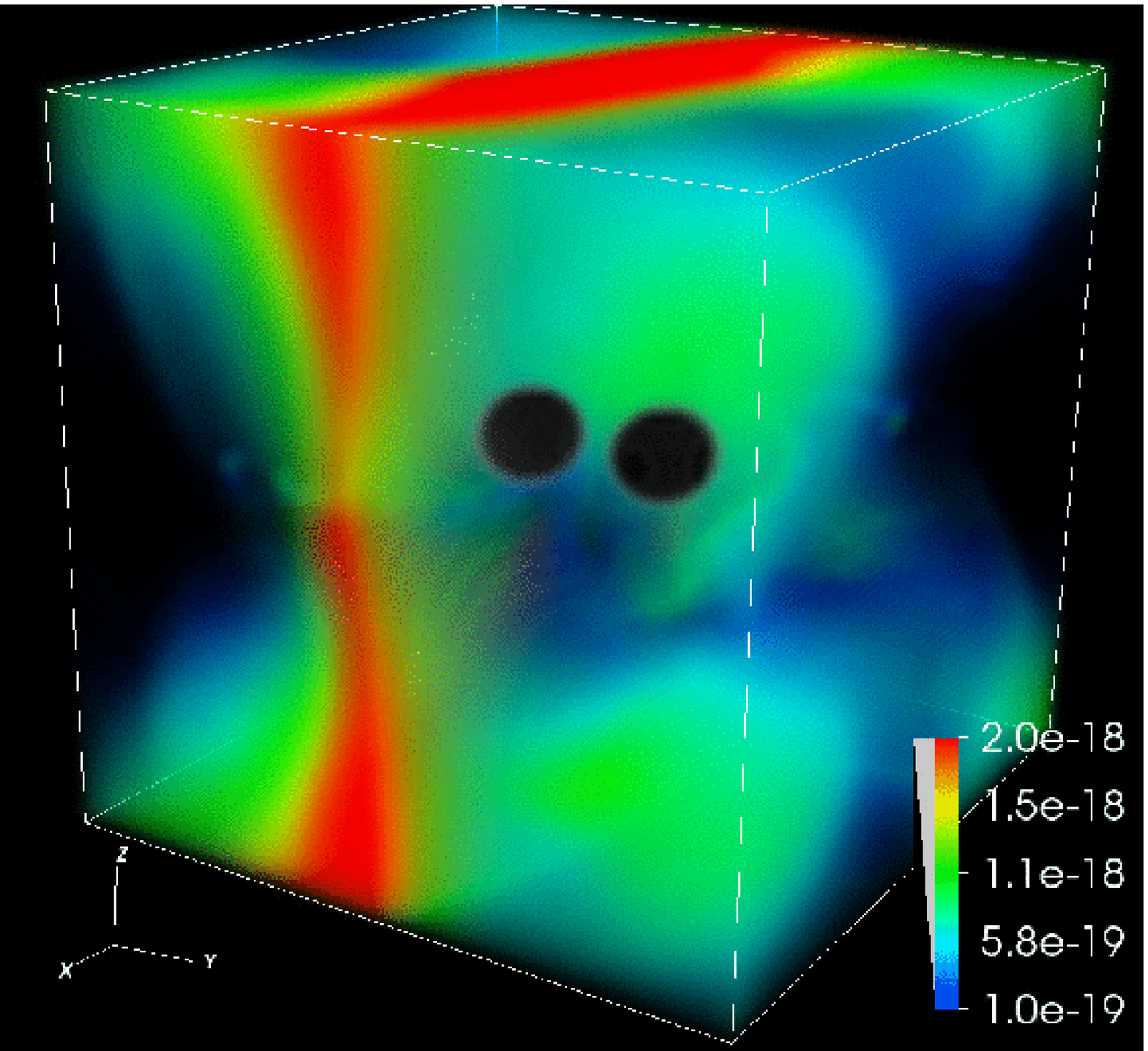, width=0.48\columnwidth}
	\epsfig{file=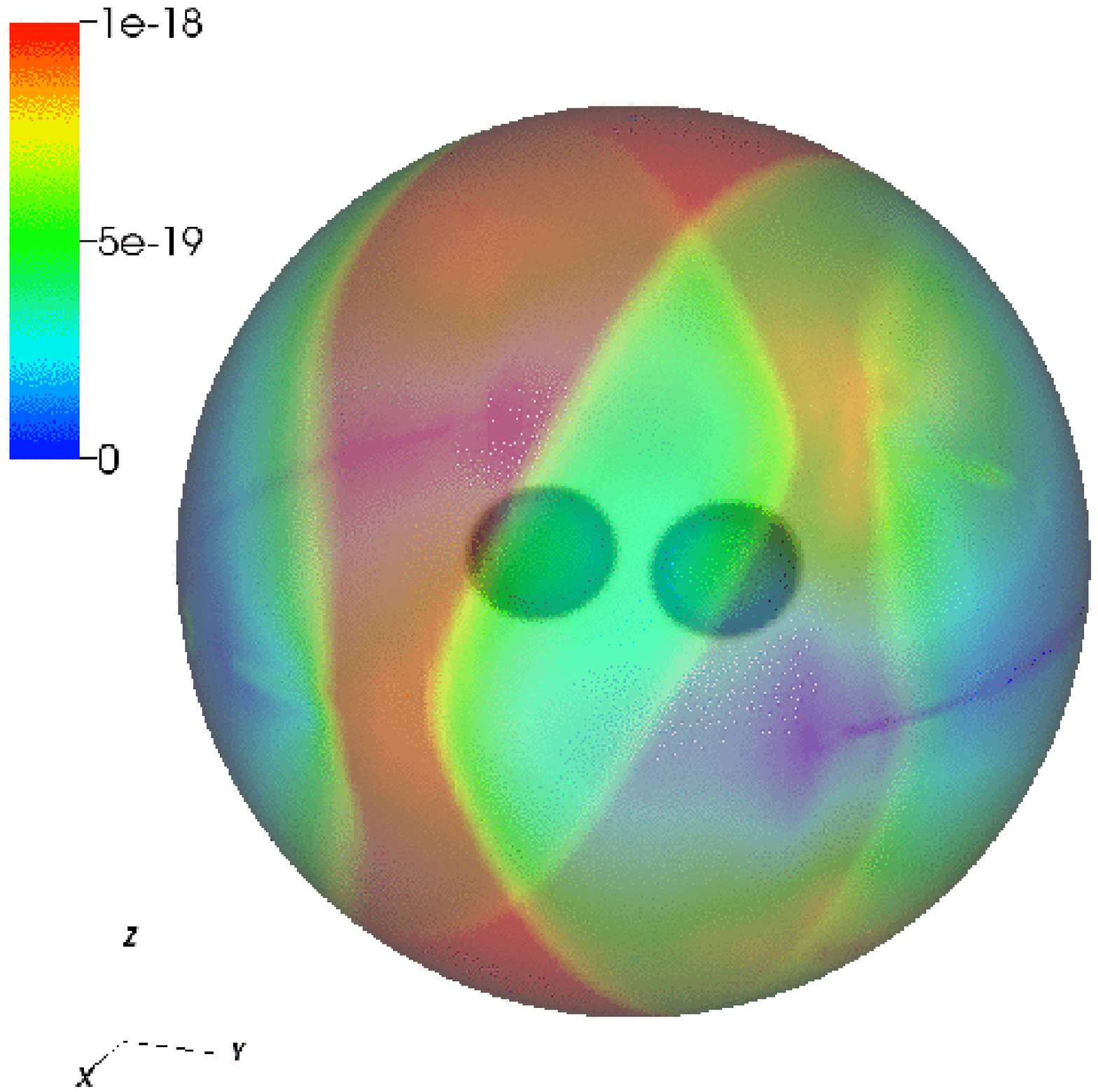, width=0.48\columnwidth} \\
	\epsfig{file=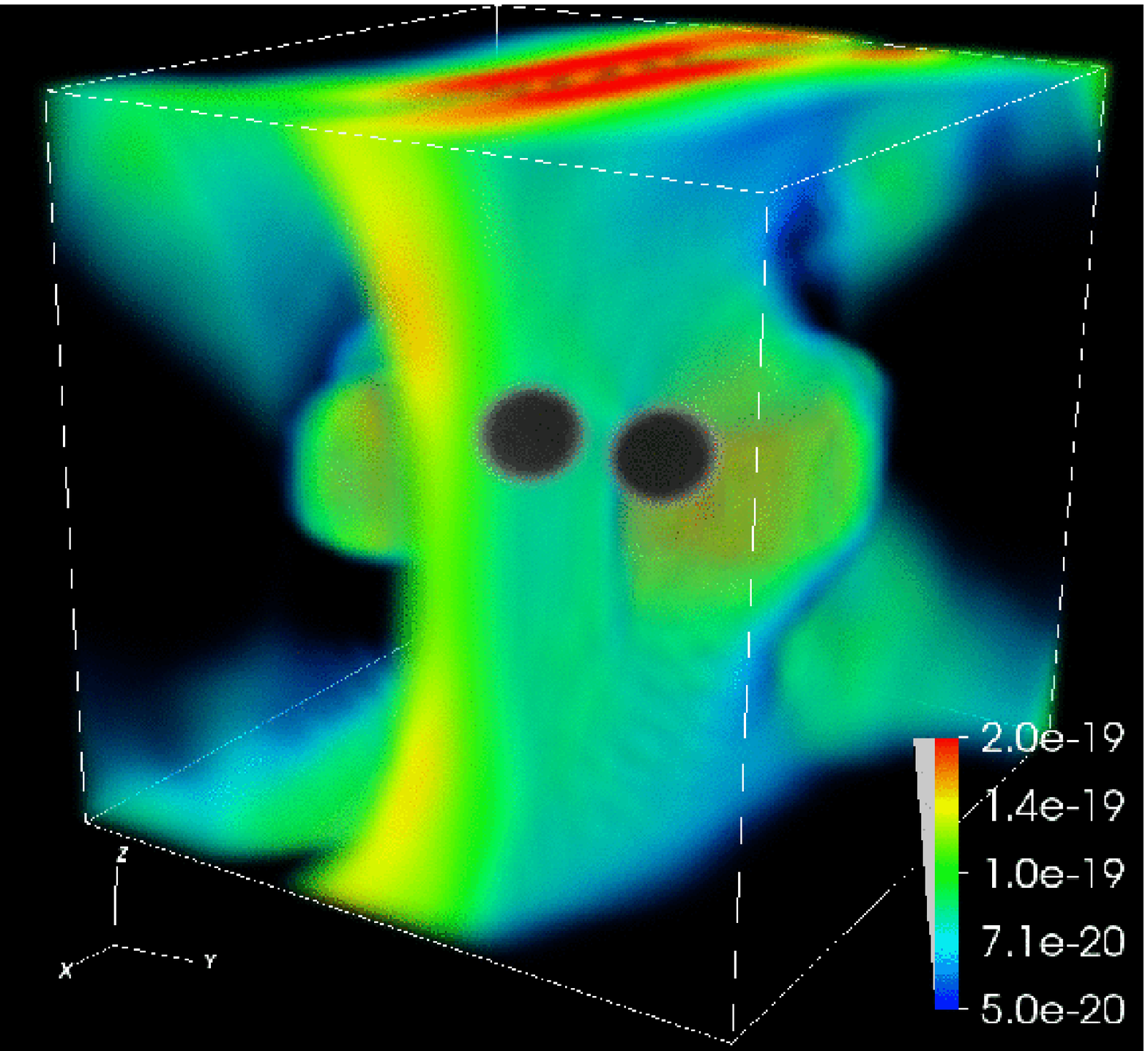, width=0.48\columnwidth}
	\epsfig{file=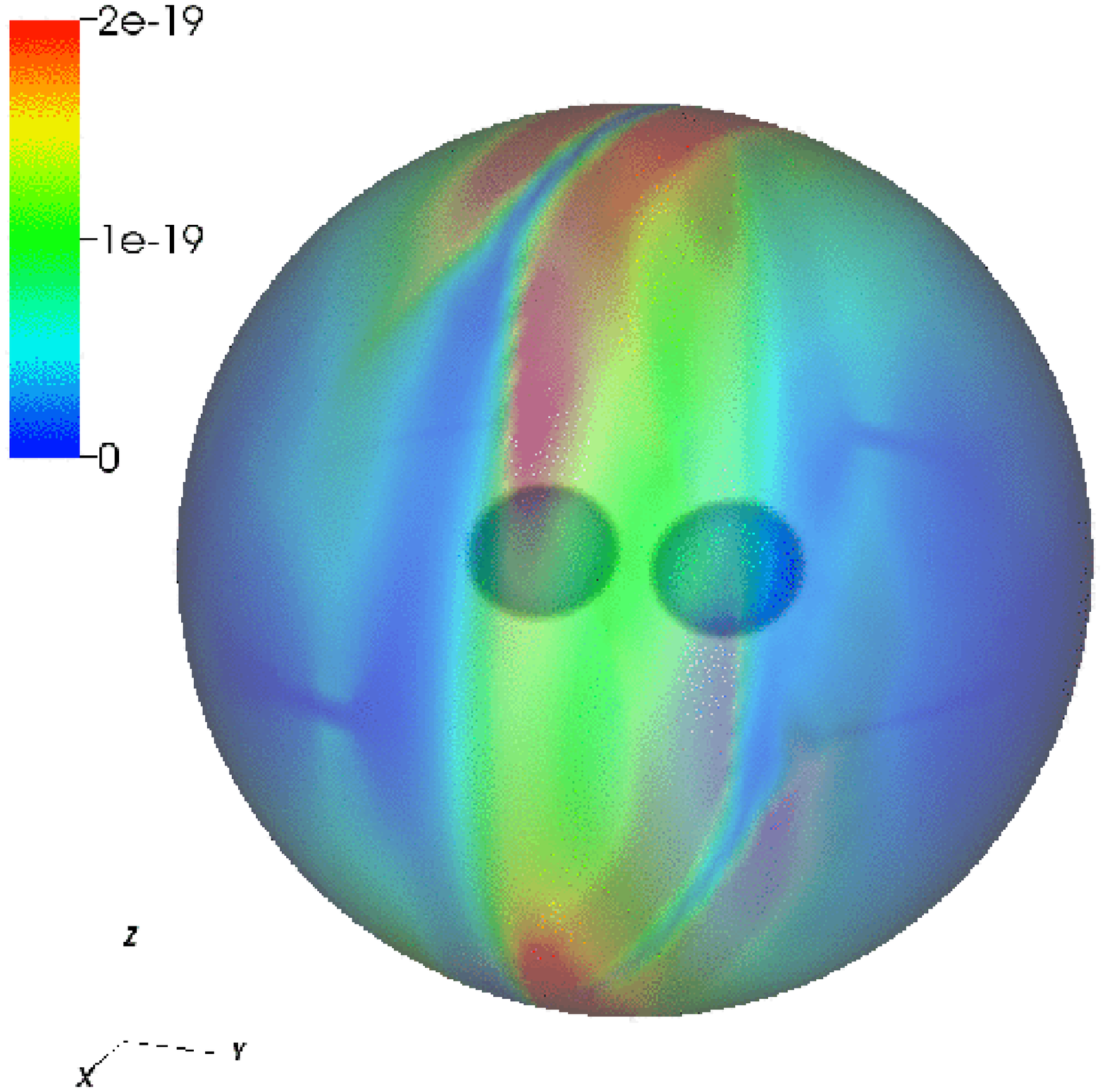, width=0.48\columnwidth} \\
	\epsfig{file=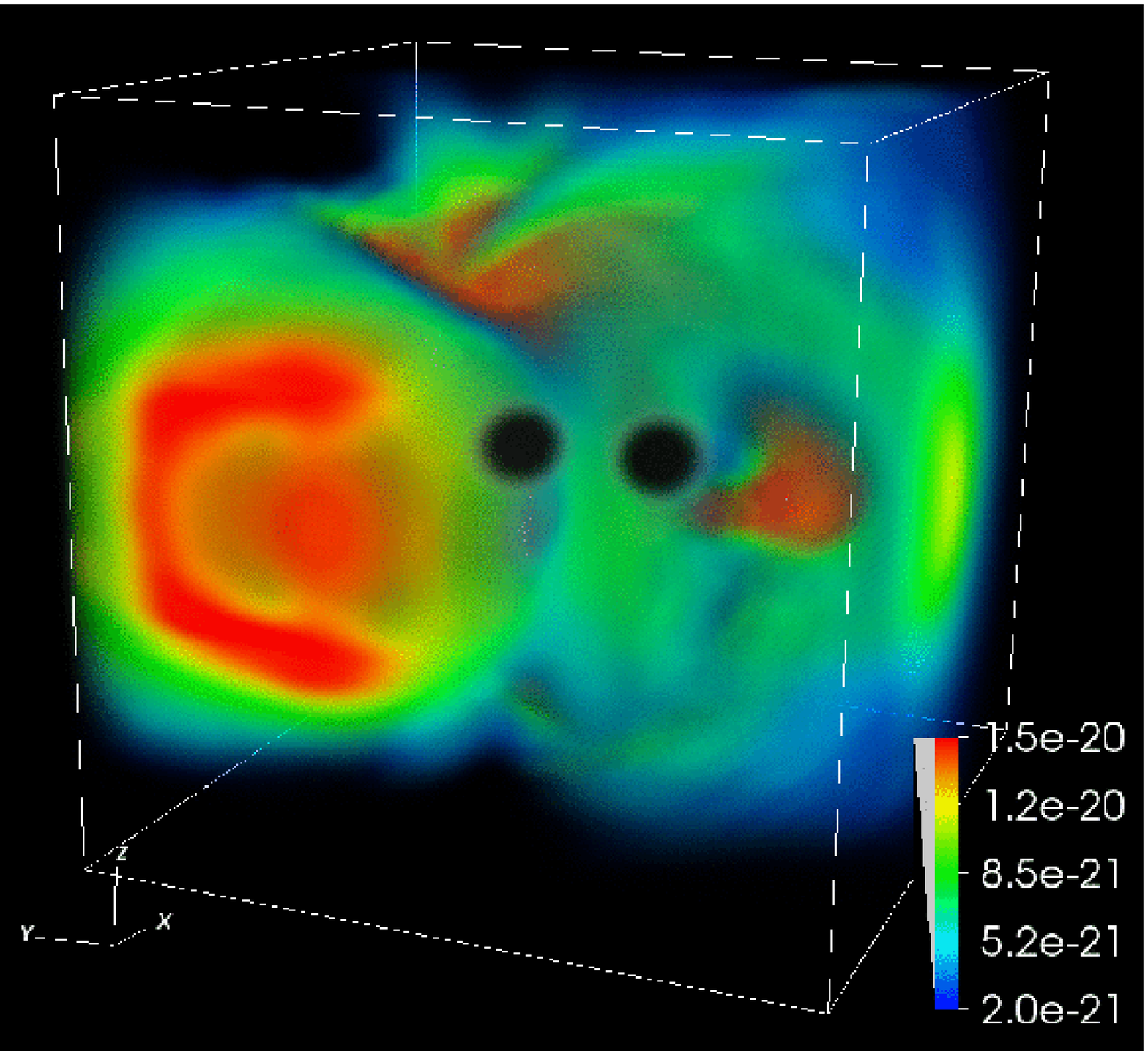, width=0.48\columnwidth}
	\epsfig{file=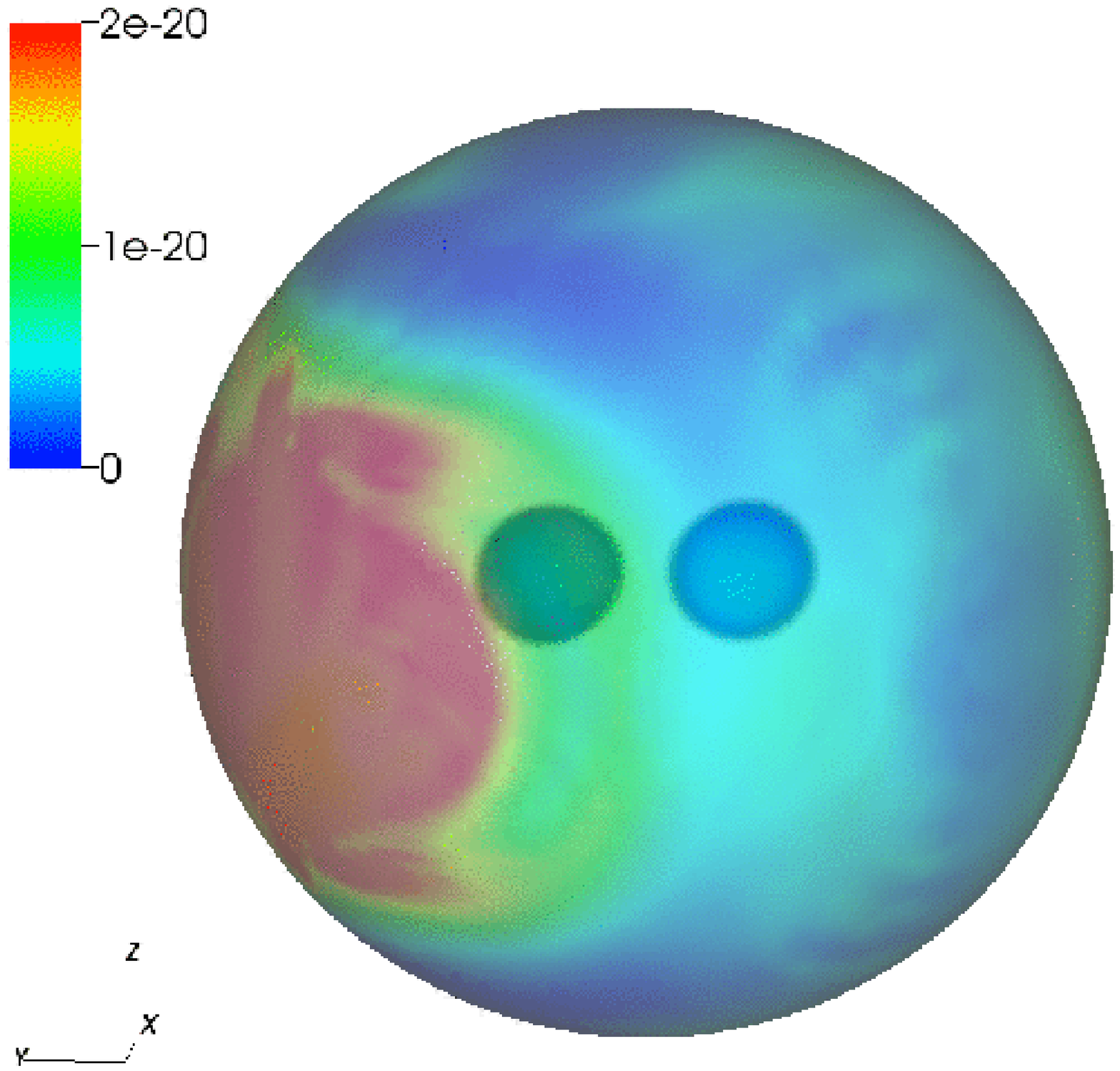, width=0.48\columnwidth}
	\caption{Poynting flux as a volume rendering (left) and evaluated on an encompassing
       sphere (right) at $t = -2.9$ ms. The {\it U/D} case (top row), the {\it U/U} case
        (middle row) and the {\it U/u} (bottom row) cases  are shown. %OLD TIME: $t = 4.1$ ms
        The sphere is located at a radius $r=80 {\rm km}$.
        The top two cases radiate more strongly away from the equatorial plane than that of
        the U/u case. 
        The U/u case radiates quite asymmetrically in the direction of the more magnetized
        star and mostly near the orbital plane. 
}
	\label{fig:sqphi2-3cases}
%\end{center}
\end{figure}

Of particular interest is the electromagnetic radiation from these configurations.
To assess their radiative properties, we study the outward Poynting flux and
find it significant (we discuss their potential observability in Section~\ref{sec:observe})
with noticeable differences among the three cases.
We show two different views of the Poynting flux for each case in Fig.~\ref{fig:sqphi2-3cases}.
On the left, a volume rendering shows the flux outside the stars. On the right, this flux is
evaluated on a binary-centered spherical surface at a radius $r=80 {\rm km}$.
As evident in the figure, both the $U/D$ and $U/U$ cases radiate strongly along the
shear layer between the two stars while the $U/u$ case does so mainly near the
equatorial plane and primarily in the direction of the strongly magnetized star.

\begin{figure}
\begin{center}
\epsfig{file=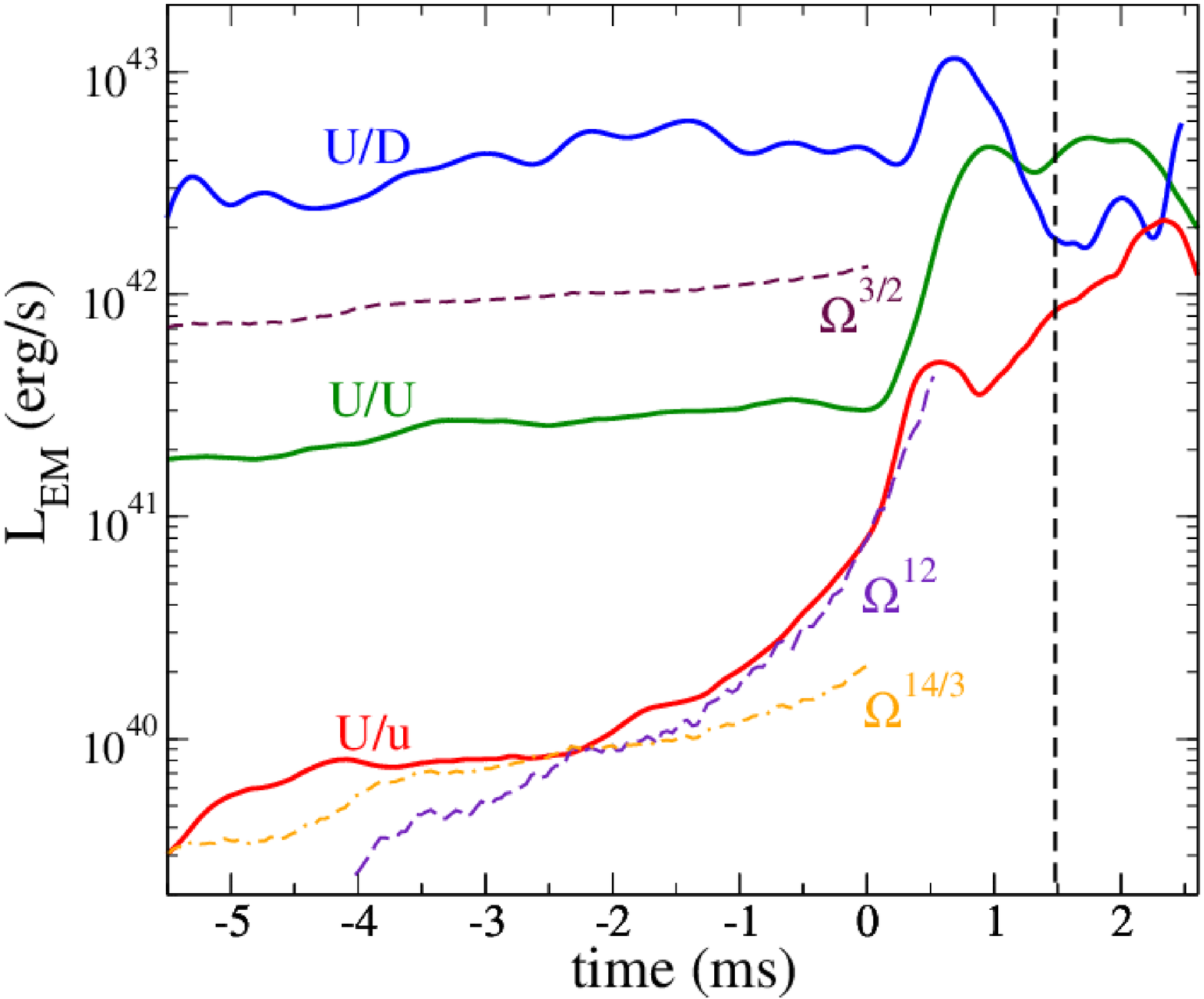,angle=-90, width=3.5in} 
\caption{
Electromagnetic luminosity for the three different magnetic field configurations.
Additionally, three curves illustrating 
$ L \propto \Omega^{p}$ with $p=\{3/2,14/3, 12\}$ are shown as guidance.
The maximum of the gravitational radiation, marked with a vertical dashed line,
occurs approximately at $t\approx 1.48$ms.
} \label{fig:energies}
\end{center}
\end{figure}

The radiation in the $U/D$ and $U/U$ cases is partially collimated. Notice that,
since the Poynting flux is symmetric across the equator ($\theta_o \rightarrow \theta_o+90^o$)
for the configurations considered here, it is sufficient to describe
only the northern hemisphere. The $U/D$ case
has a flux density in a polar cone  (with opening angle of $\theta_o<30^o$) which is
~$2.5$ times larger than the average and accounts for $1/3$ of the total radiated
energy. The $U/U$ case radiates in this polar cone about $1.9$ times larger than its average
luminosity and represents $1/4$ of the total power.
The radiation from the $U/u$ case is emitted mainly near the equatorial plane, with 
$2/3$ of the total energy radiated between $60^o<\theta_o<90^o$.
Besides the difference in angular distribution, Fig.~\ref{fig:sqphi2-3cases} indicates
that the peak $U/u$ flux is roughly a tenth that of the other two cases.
More quantitatively, we integrate the flux and display the total Poynting luminosity
for each case in Fig.~\ref{fig:energies} (solid lines). 
The behavior of these luminosities in time can be characterized in terms of powers of the orbital
frequency of the binary as a function of time, such that $L\propto \Omega^p$ 
(assuming a constant surface magnetic field).  We include on the graph a few curves for
different values of $p$ suggested by the data.

At early times, the luminosity of the $U/u$ case increases roughly
as $\Omega^{14/3}$, which is consistent with the
unipolar inductor [see Eq.~(\ref{eq:fourteenthirds})].
In contrast, the behavior of both the $U/U$ and $U/D$ cases differs from the $\Omega^{10/3}$ expectation 
of the unipolar inductor as modified for both stars being magnetized [see  Eq.~(\ref{ui_effect})].
Instead, their luminosities increase with $p\simeq 1-2$ until the
stars come into contact. Interestingly all three cases transition to much
more rapid growth ($p\approx 12$) at later times (the $U/u$ case begins this growth
a bit earlier than the other two cases).
The agreement of the slopes near merger ($t\approx 0$) for all the cases suggests that the
dynamics near merger are dominated by the formation of the hypermassive neutron
star, independent of the initial magnetic configuration.

This plot also highlights several other important details. First, the $U/D$ case is significantly
more radiative than the $U/U$ case. This disparity is interesting as the ``inner-engine''
in both cases consists of the magnetic dipole of each star and their respective orbital motion, 
which are the same except for the direction of the dipoles.
The different luminosities therefore imply a more efficient 
tapping of orbital energy with anti-aligned magnetic moments ($U/D$)  than
when they are aligned ($U/U$), possibly due to the 
additional energy radiated by the release of magnetic tension in the $U/D$ case through
reconnections near the stars.
We have monitored that the  electromagnetic energy in the interior of the stars remains
essentially constant during the inspiral, with a sudden increase when the star
surfaces touch.

Of course, as the stars merge the shear between them will considerably increase
the magnetic energy through conversion of mechanical energy (e.g. through
Kelvin-Helmholtz instabilities~\cite{Price:2006fi,2008PhRvL.100s1101A}).
Although an exponential growth of the magnetic energy has been observed
in local simulations~\cite{2009ApJ...692L..40Z,2010A&A...515A..30O}, global simulations of binary NSs (i.e., like ours)
have not yet reached the required accuracy to capture the dynamics
occurring at the smallest scales. As a result, there is only moderate growth
of the magnetic energy, which saturates at values lower than in local simulations.

Nevertheless, there are some useful qualitative
observations from our simulations beyond the merger epoch.
First, the magnetic flux through a hemisphere around the individual
stars decreases after the merger; due to the reconnection of magnetic field lines,
part of the ordered dipolar magnetosphere is ejected soon after the formation of
the rotating hypermassive neutron star. 
As the stars lose their ordered dipolar magnetospheres and magnetic flux, the
luminosity is expected to decrease (although this behavior might change 
due to the increase in the magnetic field strength at the merger).
Notice however that, as discussed in~\cite{Lehner:2011aa}, an ordered
magnetosphere may emerge again at later times by dynamo action in a surface
shear layer. 
Second, the luminosity of the $U/u$ case grows even after
merger until it becomes comparable to that of the other two cases. This growth is
expected because the final configuration for all our cases is always an aligned
rotator of the same rotational velocity and a magnetic dipole moment of comparable
strength and extent.

%---------------------------------------------
%\begin{widetext}
%\end{widetext}
%---------------------------------------------

%---------------------------------------------
%\begin{widetext}
%\end{widetext}
%---------------------------------------------

We can also consider this system as if the stars were immersed in electrovacuum instead
of being surrounded by tenuous plasma (i.e. magnetospheres).
Recall that orbiting dipoles with equal, aligned moments produce no
electromagnetic radiation at dipole order, while anti-aligned moments do (see Appendix~\ref{appendixA}). 
Thus that the $U/D$ case radiates more is not surprising. Nevertheless, such an argument
resorting to electromagnetism would suggest that the $U/u$ case should be more radiative than
the $U/U$ case because the effective dipole of the $U/u$ is non-zero.  However, in our calculations, it is instead
the aligned case that is much more radiative, and the failure here
indicates that this electrovacuum analogy 
can be taken only so far.

Another argument that ignores the magnetosphere suggests
that the anti-aligned dipoles {\em liberate} potential energy as they get closer, whereas the aligned dipoles
{\em require} the input of electromagnetic potential energy as they approach.
One problem with this argument that it would predict the radiation of the $U/u$ case
to fall between the other two, which is not the case.

%---------------------------------------------
%\begin{widetext}
%\end{widetext}
%---------------------------------------------

It is interesting to point out here that in the $U/u$ case the amount
of energy dissipated as Joule heating $J_i E^i$ is comparable to the Poynting energy radiated.
For the $U/D$ and $U/U$ cases, on the other hand, the energy dissipated
as heat is only roughly a third  of the radiated one.
In summary, the Poynting integration indicates that magnetosphere
interactions in the $U/D$ and $U/U$ cases yield additional radiation, and with
a different orbital dependence,  than what could be explained via a simple
unipolar induction model. This result is reinforced by the important
differences in the Joule heat observed between the $U/u$ case and the
others.

%%%%%%%%%%%%%%%%%%%%%%%%%%%%%%%%%%%%%%%%%%%%%
\section{Observability prospects}
\label{sec:observe}
%%%%%%%%%%%%%%%%%%%%%%%%%%%%%%%%%%%%%%%%%%%%%
In this work we have studied the basic phenomenology of magnetosphere
behavior of an inspiraling, magnetized binary neutron star system.
Our results imply that the late orbiting stages of such a system can induce
strong electromagnetic emissions, in addition to strong gravitational wave output.
These electromagnetic emissions are sourced by the ability of the magnetosphere
to tap rotational energy from the system and to induce both current sheets and Poynting flux.
Interestingly, current sheets
in the system are heterogeneous and have an imprint of the orbital behavior which
could give time-varying signals that aid in detecting these systems
electromagnetically. While a detailed analysis of these signatures is outside the scope of this work, we
comment here on some possible relevant options.

%%%%%%%%%%%%%%%%%%%%%%%%%%%%%%%%%%%%%%%%%%%%%
\subsection{Thermal spectra}

The combination of very high energy density with a strong magnetic field
as studied here
is expected to produce an optically thick environment emitting roughly
as a black-body (see Appendix~\ref{appendixB} for a more detailed discussion).
The effective temperature of this radiation can be calculated by balancing the Poynting
luminosity (absorbed by the magnetospheric plasma) 
with the (black-body) radiation emitted by the magnetosphere.
Further, recall that the magnetic field will not affect the pre-merger dynamics 
provided that the strengths considered are $< 10^{17}$G. Therefore, the Poynting luminosities used
to estimate the temperature can be rescaled 
as ${\cal L} \propto B^2$. We can then arrive at an expression for the temperature by
extracting the total Poynting
luminosity ${\cal L}$ emitted within a radius $r_o \approx 30 {\rm km}$ (i.e., around
$t \approx -5 {\rm ms}$) as
\begin{equation}\label{blackbody_temperature}
     T = \left( 10^7 {\rm K} \right) S_{AB} \left(B_{11}\right)^{1/2}
\end{equation}
where $B_{11}=B/10^{11}$G and $S_{AB}$ characterizes the configuration
of the initial magnetic field.
In particular, we find
$S_{U/U} = 4.7$, $S_{U/u} = 2$, and $S_{U/D} = 10$ for the three cases studied here.

Notice that the luminosity in the $U/u$ case obtained from our simulations
has the same order of magnitude as the simple estimate
obtained assuming the unipolar inductor model for the strongly magnetized star
with a weakly magnetized companion in Eq.~(\ref{Hansen_Lyutikov3}). 
This agreement leads to similar effective temperatures. 
From the effective temperature, 
computation of  the peak frequency of
the black-body radiation (i.e., via Wien's displacement law
$\nu_{\rm peak} = \left( 5.88 \times 10^{10} {\rm Hz}/{\rm K} \right) T$), falls within the
hard X-ray range. 
Notice that in the extreme case where the primary star (for the $U/u$ case)
is a magnetar with $B\sim 10^{15}$G the luminosity will increase 
to ${\cal L}_{U/u} \approx 1.6 \times 10^{47} {\rm erg/s}$ with
a black-body temperature $T_{U/u} = 2 \times 10^9$ K,
well inside the $\gamma$-ray range.

%%%%%%%%%%%%%%%%%%%%%%%%%%%%%%%%%%%%%%%%%%%%%%%%%%%%
\subsection{Non-thermal components}

Neutron stars and their magnetic configurations are responsible for a number of
non-thermal emissions such as pulsars and soft gamma-ray repeaters, and clearly
the systems studied here will have non-thermal emissions. 
One possibility, as argued in~\cite{2012arXiv1212.0333M}, is that the system's strong Poynting flux
drives a relativistic bubble that pushes into the surrounding inter-stellar medium.
At sufficiently large  separations, the shock at the interface of the bubble becomes collisional and can be 
responsible for synchrotron radio emission. The details of the shock behavior however is more involved
than those worked-out in~\cite{2012arXiv1212.0333M}. As our studies indicate, the Poynting flux is 
asymmetric and its complex time dependence is intimately tied to the initial field configuration of the stars.

\begin{figure}
\begin{center}
\epsfig{file=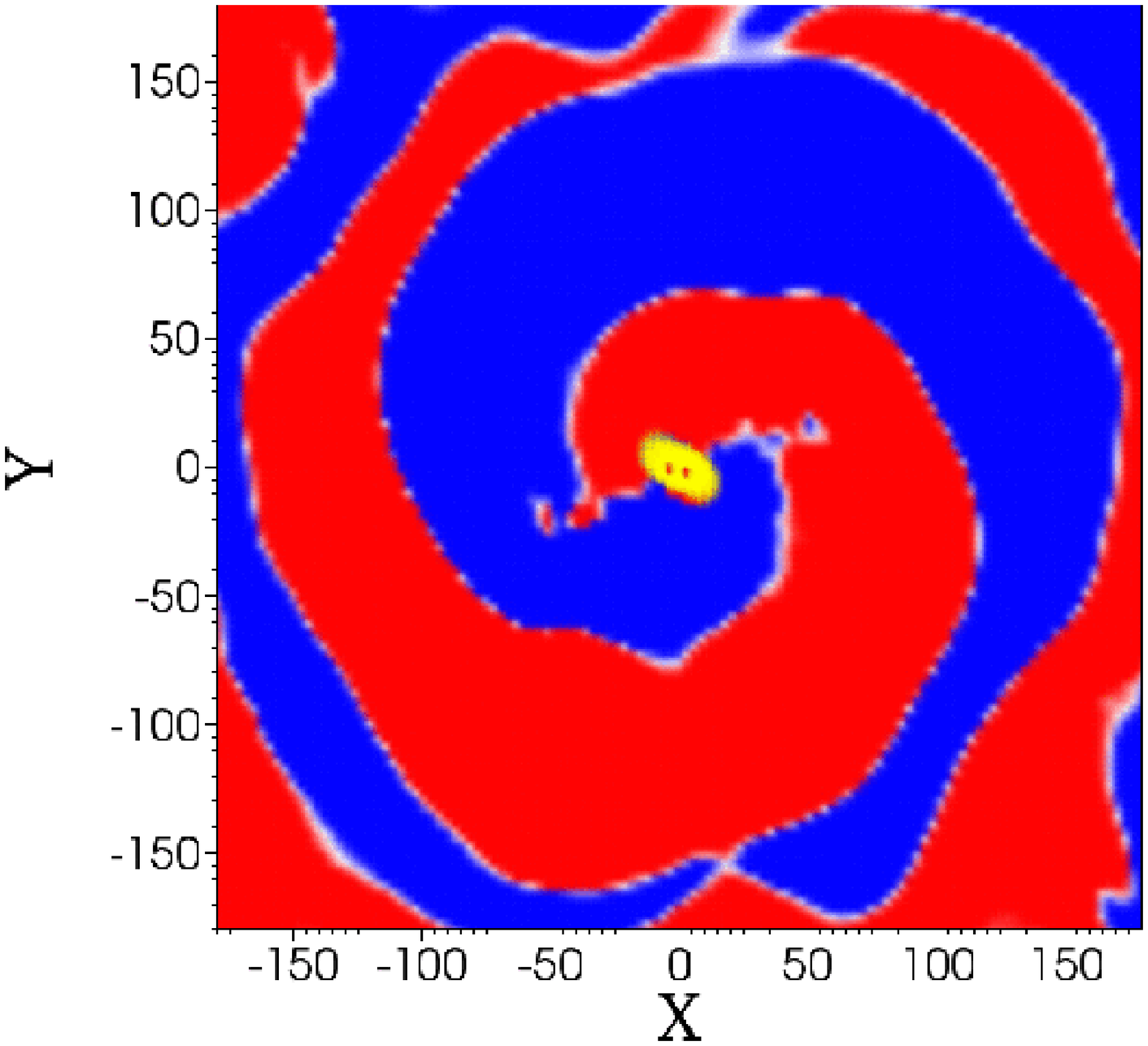, width=0.85\columnwidth}
\epsfig{file=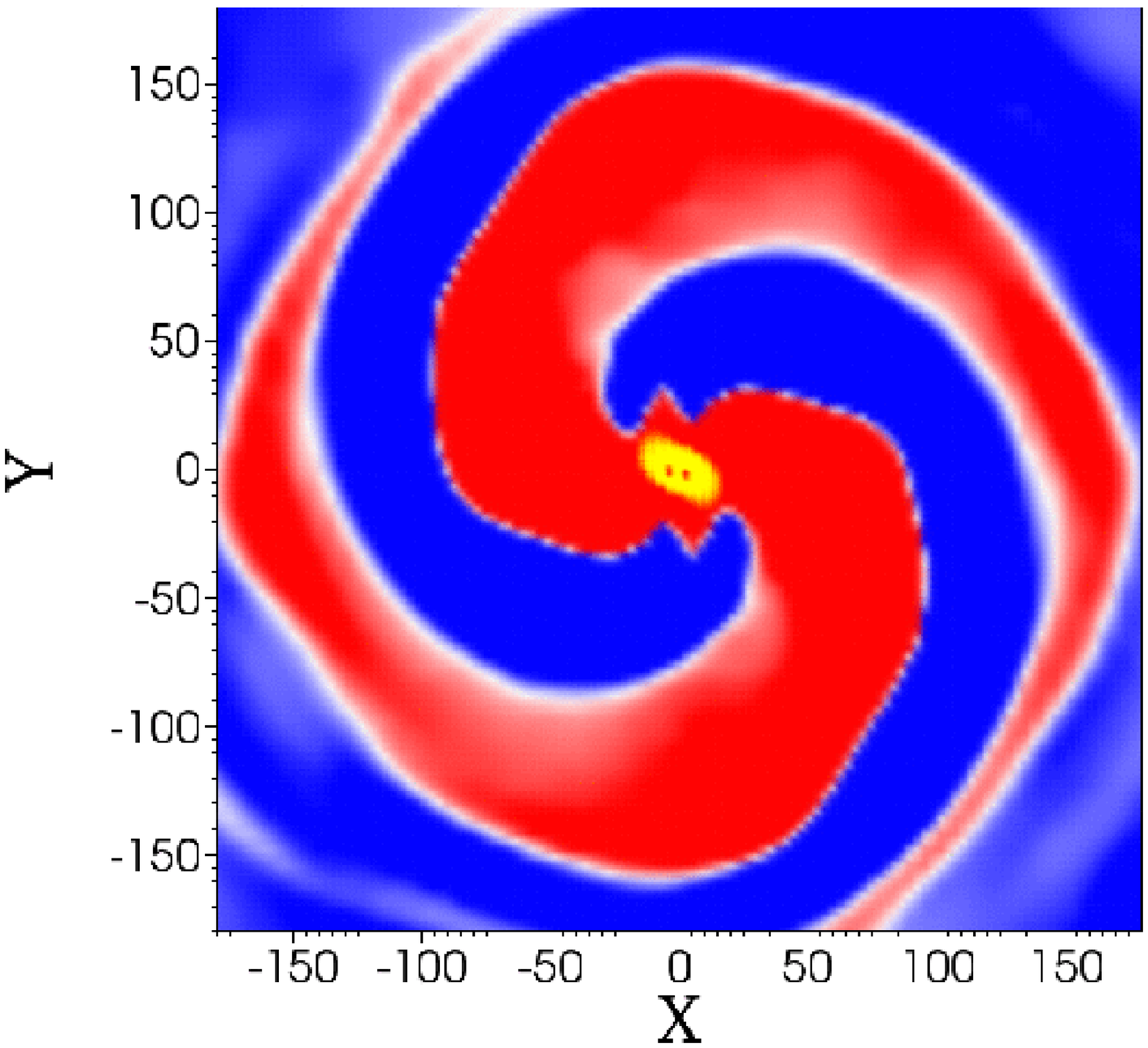, width=0.85\columnwidth}
\epsfig{file=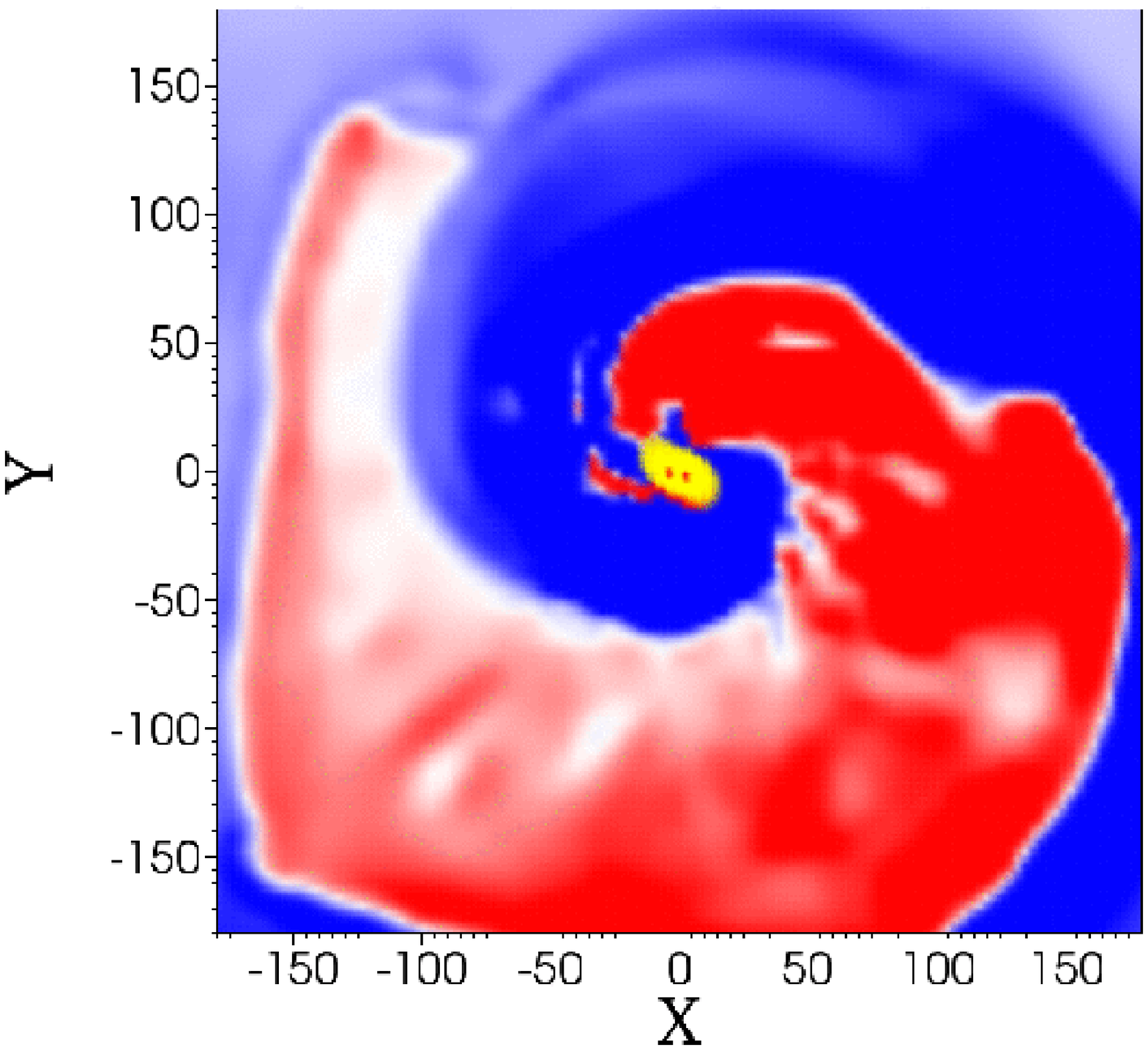, width=0.85\columnwidth}
\caption{Polarity (red positive, blue negative) of the $B_z$ component
of the magnetic field at $z=+7.4$km above the orbital plane
at late times ($t \approx 1.4$ms; after the stars have already touched)
for the $U/D$, $U/U$ and $U/u$ cases respectively.
The pattern resembles a  ``striped'' pattern.  
The density of the merging binary is also shown at the center of the figures.
} \label{fig:stripes}
\end{center}
\end{figure}

As the orbit tightens, the shock would become collisionless and different processes would become important. %come into play.
In particular, the magnetic field far from the stars resembles 
that of the striped winds from pulsars. As shown in Fig.~\ref{fig:stripes}, the orbiting
behavior induces an outflowing wind with regions of opposite magnetic field polarity similar
to those discussed in the context of oblique rotators~\cite{2011ApJ...741...39S}.
Models arguing for strong 
particle acceleration in pulsar winds~\cite{2011ApJ...741...39S} through shock-driven magnetic reconnections would consequently also be applicable here.

Furthermore, accelerating fields can arise naturally at gaps
(\cite{1995ApJ...438..314R,2000ApJ...537..964C,Muslimov:2003yz,Muslimov:2004vj}) energizing
a population of particles that emit high energy, synchrotron radiation.
Another emission mechanism as argued in e.g~\cite{2011SSRv..160...45U,Uzdensky:2012tf} asserts that
strong cooling at the current sheets, such as we see in the $U/U$ and $U/D$ cases, can give rise to  gamma-rays 
via synchrotron radiation~\cite{Uzdensky:2012tf} or inverse Compton scattering (see for instance~\cite{2012arXiv1208.5329L}).
Flares of intense gamma-rays at the radiation-reaction limit 
have been observed from
the Crab Nebula but without significantly detected components in the 
rest of the electromagnetic spectrum~\cite{2012arXiv1211.3997W,2012ApJ...749...26B}.
These observations perhaps implicate regions of magnetic reconnection as
sites of tremendous particle acceleration.
Here we are in a situation with a very dynamic magnetic configuration
and favorable sites for particle acceleration likely arise.

\section{Final words}
We have shown that global magnetic fields within a binary system can give
rise to a rich phenomenology that powers strong emissions on the electromagnetic
side ($\simeq 10^{40-43} (B/10^{11}{\rm G})^2$ erg/s) prior to the merger. These luminosities
are at the level of the brightest pulsars and beyond and would bear
particular characteristics tied to the orbital behavior. Such high luminosities, together with the power
emitted in gravitational waves, indicate that the system is strongly radiative in multiple bands
and channels. We have also identified possible features that can lead to observable
signals tied to the orbital behavior of the system. The consequential time-dependent
nature of possible emissions might help in their detection, especially if some prior 
localization (in time and space) is provided by gravitational wave information.
The details of the emission mechanism however are still uncertain.  Different emission mechanisms are expected near
the current sheets, where strong cooling can give rise to  gamma-rays~\cite{2011SSRv..160...45U,Uzdensky:2012tf} produced
via either synchrotron~\cite{Uzdensky:2012tf} or inverse Compton scattering~\cite{2012arXiv1208.5329L} (see also 
discussion in~\cite{1996A&AS..120C..49A}).  Also, at gaps, accelerating fields can develop~\cite{1995ApJ...438..314R,2000ApJ...537..964C,Muslimov:2003yz,Muslimov:2004vj} and energize
particles which could also emit at high energies via curvature and synchrotron 
radiation. Understanding which of these mechanisms are the most relevant 
is yet unknown even in pulsar models so there is a large degree of uncertainty in this question.
At a simple level however,
a relativistically expanding electron-positron wind sourced by energy dissipation
and magnetohydrodynamical waves in between the stars could create an X-ray 
signature~\cite{2001MNRAS.322..695H} preceding or coincident with the merger.
Thus, ISS-Lobster~\cite{Camp:2013cwa} with its high sensitivity and wide field of view
would be very well suited for detecting the associated
electromagnetic counterpart to a binary neutron star merger. For the values calculated
here these sources would be observable to distances of $10^{0 - 2} (B/10^{11}G)$~Mpc 
(assuming 10\% efficient conversion of Poynting flux).

%%%%%%%%%%%%%%%%%%%%%%%%%%%%%%%%%%%%%%%%%%%%%%%%%%%%%%%%%%%%%%%%%%%%
%
%   A C K N O W L E D G M E N T S
%
%%%%%%%%%%%%%%%%%%%%%%%%%%%%%%%%%%%%%%%%%%%%%%%%%%%%%%%%%%%%%%%%%%%%
\vspace{0.5cm}

\noindent{\bf{\em Acknowledgments:}}
It is a pleasure to thank A. Broderick, J. McKinney, A. Spitkovksy and 
C. Thompson for interesting discussions.
This work was supported by the NSF under grants PHY-0969827~(LIU), 
PHY-0969811~(BYU), NASA's ATP program through grant NNX13AH01G,
NSERC through a Discovery Grant (to LL) and CIFAR (to LL). 
C.P acknowledges support by the Jeffrey L.~Bishop Fellowship.
Research at Perimeter
Institute is supported through Industry Canada and by the Province of Ontario
through the Ministry of Research \& Innovation.  Computations were
performed at XSEDE and Scinet.

%%%%%%%%%%%%%%%%%%%%%%%%%%%%%%%%%%%%%%%%%%%%%%%%%%%%%%%%%%%%%%%%%%%%%%%
\appendix
\section{Electrovacuum calculation of radiation}\label{appendixA}
%%%%%%%%%%%%%%%%%%%%%%%%%%%%%%%%%%%%%%%%%%%%%%%%%%%%%%%%%%%%%%%%%%%%%%%
\renewcommand{\theequation}{A-\arabic{equation}}
\setcounter{equation}{0}
%%%%%%%%%%%%%%%%%%%%%%%%%%%%%%%%%%%%%%%%%%%%%%%

The radiated electromagnetic energy of two dipoles in electrovacuum can be
obtained via the Post-Newtonian equations of motion. The rate of energy loss due to electro-magnetic radiation
to 2.5PN order (which includes gravitational radiation effects) has been
presented in~\cite{Ioka:2000yb,Blanchet} under the assumption that the stars' magnetic dipoles 
remain constant and neglecting spin-orbit effects, 
\begin{equation}
	\label{eqn:dipdip-radn-PN}
	\begin{array}{lcl}
	\frac{dE}{dt}	&=& 
		-\frac{2}{15} \frac{m^2}{r^6} \left[ 2\mu^2_{\rm eff} \left\{ v^2 - 6\dot{r} \left( \hat{n}\cdot\vec{v}\right) + 9\dot{r}^2 \right\} \right.		\\
			&&	\left. - 
		\left\{ \vec{\mu}_{\rm eff} \cdot \left( \vec{v} - 2\dot{r}\hat{n} \right) \right\}^2 \right]
	\end{array}
\end{equation}
where $r = |\vec{r}_2-\vec{r}_1|$ is the separation between the stars,
$\vec{v} = \vec{v}_2-\vec{v}_1$ is the relative velocity,
and $\hat{n} = \vec{r}/{r}$ represents the unit vector between the stars in the center of mass frame;
$m = m_1 + m_2$ is the total mass of the system; and
$\vec{\mu}_{\rm eff} = \left( m_2 \vec{\mu}_{(1)} - m_1 \vec{\mu}_{(2)} \right)/m$
represents the effective magnetic dipole of the system.
In the case
of an equal mass binary system with identical magnetic dipoles
(i.e. same direction and magnitude; $U/U$) $\vec{\mu}_{\rm eff} =0$ and no radiation is produced at
this order. On the other hand, the choice of anti-aligned moments ($U/D$) maximizes the predicted radiation.

Moreover, assuming a circular orbit with $\dot{r}=0$, an effective
magnetic dipole moment perpendicular to the velocity, 
and a Keplerian rotational frequency $\Omega$, it is straightforward to determine
that the radiated energy scales as $\Omega^{14/3}$,
at leading order.
Notice that this scaling is the same as in Eq.~(\ref{eq:fourteenthirds}), the 
estimate provided by the unipolar inductor
model for the $U/u$ case, although its
magnitude in the electrovacuum case is a few orders of magnitude smaller
than in the force-free case.

%%%%%%%%%%%%%%%%%%%%%%%%%%%%%%%%%%%%%%%%%%%%%%
\renewcommand{\theequation}{B-\arabic{equation}}
\setcounter{equation}{0}
%%%%%%%%%%%%%%%%%%%%%%%%%%%%%%%%%%%%%%%%%%%%%%%%%%%%%%%%%%%%%%%%%%%%%%%
\section{Relativistic outflow and black-body radiation}\label{appendixB}
%%%%%%%%%%%%%%%%%%%%%%%%%%%%%%%%%%%%%%%%%%%%%%%%%%%%%%%%%%%%%%%%%%%%%%%

As has been noted in~\cite{2011SSRv..160...45U}, the release of 
electromagnetic energy with sufficiently high energy density will 
produce a relativistic outflow of electron-positron pair plasma with
a roughly black-body spectrum as long as the medium is optically thick.
Such a condition is naturally 
induced by
a magnetic field with strength on the order
of the quantum critical field $B_{\rm QED}=4.4 \times 10^{13} G$. At this strength,
the magnetic energy density is high enough for copious pair-production,
and the resulting medium becomes optically thick due to electron scattering
and pair production.
In this regime, the radiated energy will produce a relativistic wind of
radiation and pairs~\cite{1986ApJ...308L..43P}. % which, in the spherically

To estimate the effective temperature 
we can proceed as follows.
First, let us define the normalized temperature $\Theta_e \equiv k_B T/(m_e c^2)$
and the normalized magnetic field $b \equiv B/B_{\rm QED}$ and recall that
pair-production can lead to an optically thick regime (for which it is safe to
assume approximate black-body radiation). 
In such a regime, we can assume 
that in equilibrium the electromagnetic energy injection must equal
the total pressure leading to the
pressure-balance relation
\begin{eqnarray}\label{balance_temp}
   P_{\rm rad} + P_{\rm pairs} + P_{\rm baryon} = \frac{(b B_{\rm QED})^2}{8 \pi}
\end{eqnarray}
where $P_{\rm rad},P_{\rm pairs},P_{\rm baryon}$ are the pressures associated with
photons, electro-positron pairs, and baryons respectively. 
Next, we set $P_{\rm rad}=a T^4/3$ and assume that the baryon pressure of the 
charge  density is small. Furthermore, we note that the resulting pressure of
electron-positron pairs in thermal equilibrium is negligible for
$\Theta_e << 1$, while for $\Theta_e \ge 1$ it is comparable to
the radiation pressure $P_{\rm pairs}=7 P_{\rm rad}/4$~\cite{2011SSRv..160...45U}.
Combining these assumptions with Eq.~(\ref{balance_temp}) yields
\begin{eqnarray}\label{effective_temp}
   \Theta_e \sim \kappa \, b^{1/2}
\end{eqnarray}
where $\kappa$ is just a numerical coefficient equal either to 
$2.2$ for $\Theta_e \ll 1$ or $1.7$ for $\Theta_e \ge 1$.
In the following we consider an intermediate value
$\Theta_e \sim 2\, b^{1/2}$ which will be roughly valid for any temperature. 

Now, as discussed for large magnetic fields $b>1$ (i.e., corresponding to high
effective temperatures $\Theta_e \geq 1$) a dense pair-plasma is easily
produced and the medium is optically thick. 
For low temperatures, $\Theta_e <1$, the density of pairs is  
$n_{\pm} = \left(4.4 \times 10^{30} {\rm cm}^{-3}\right) \Theta_e^{3/2} e^{-1/\Theta_e}$,
leading to an opacity $\kappa \sim (\sigma_T n_{\pm}) \sim
\left(2.9 \times 10^{6} {\rm cm}^{-1}\right) \Theta_e^{3/2} e^{-1/\Theta_e}$.
Taking, for instance, a realistic magnetic field value of $B \sim 4 \times 10^{11}$G
yields a normalized temperature of
$\Theta_e \sim 0.2$. This choice also gives an optical depth 
$\lambda \sim \kappa^{-1} \simeq 6 \times 10^{-4} {\rm cm}$ and  
a characteristic diffusion time $\tau_{\rm diff} \sim R^2/(\lambda c)$ (in
terms of some characteristic length scale $R$). 
A lower bound for $\tau_{\rm diff}$ is obtained by choosing
$R \sim 10^6 {\rm cm}$ (commensurate with a typical NS radius) with
$\tau_{\rm diff} \ge 6 \times 10^4 {\rm s}$. % for a typical size of a neutron star.

These arguments suggest that even for more moderate field strengths below $B_{\rm QED}$
but above $\simeq 10^{11}$G, the system will become optically thick to photons and radiate
with a black-body spectrum.
The characteristic temperature of the radiation can be determined by equating
the black-body power given by the Stefan-Boltzmann law with the sum of
the Poynting flux and dissipated power
\begin{eqnarray}\label{balance_energy}
   \sigma_{\rm SB} T^4 \sim S_{\rm Poynting} + S_{\rm diss}
                   \sim \frac{ {\cal L} }{4 \pi r_o^2} \, .
\end{eqnarray}
Here, ${\cal L}$ is the total power radiated either as Poynting
flux or dissipated at the shear layer, with 
$r_o$ the characteristic radius where most of this energy is injected.
Using Eq.~(\ref{Hansen_Lyutikov3}), setting $r_o=a$, and assuming
that the dissipated energy at shear layers is a small fraction of the
Poynting flux,
we obtain
\begin{eqnarray}\label{balance_energy_theoretical}
%   \sigma_{\rm SB} T^4 &\sim& 1.5 \times 10^{26}\, \left( \frac{B_*}{10^{11}G} \right)^2 
%                \, \left( \frac{a}{30 {\rm km}} \right)^{-9} {\rm ergs/(cm^2 s)}
   \sigma_{\rm SB} T^4 &\sim& \left( 1.5 \times 10^{26} {\rm ergs/(cm^2 s)} \right) \, \left( \frac{B_*}{10^{11}G} \right)^2 
                \, \left( \frac{a}{30 {\rm km}} \right)^{-9}
\end{eqnarray}
leading to a very high effective temperature
\begin{eqnarray}\label{temperature_theoretical}
   T &\sim& \left( 4 \times  10^{7} {\rm K}\right) \left( \frac{B_*}{10^{11}G} \right)^{1/2} 
                \, \left( \frac{a}{30 {\rm  km}} \right)^{-9/4}.
\end{eqnarray}
However, as mentioned, the applicability of this estimate depends strongly
on the magnetic field behavior and so on the stage at which it can be adopted.

%%%%%%%%%%%%%%%%%%%%%%%%%%%%%%%%%%%%%%%%%%%%%%%%%%%%%%%%%%%%%%%%%%%%%%%%%%%%%%%%%%
%\bibliographystyle{ieeetr}
% this provides more info and also links to papers which is convenient:
\bibliographystyle{utphys}
\bibliography{./passiveBNS}
\end{document}